\documentclass[reprint,a4paper,australian,superscriptaddress,amsmath,aps,prd,twocolumn,noshowpacs,floatfix,preprintnumbers]{revtex4-1}

\usepackage{amsmath}
\usepackage{enumerate}
\usepackage{amssymb}
\usepackage{amsfonts}
\usepackage{graphicx}
\usepackage{bm}
\usepackage{subfigure}
\usepackage{amscd}
\usepackage{epsfig}
\usepackage{hhline,multirow}
\usepackage{dcolumn}
\usepackage{url}
\usepackage[dvipsnames]{xcolor}
\definecolor{DarkBlue}{rgb}{0.0, 0.0, 0.4}
\definecolor{darkgreen}{cmyk}{1,0,1,0.5}

\usepackage[colorlinks=true,linkcolor=blue,urlcolor=DarkBlue]{hyperref}
\usepackage{indentfirst}
\usepackage{psfrag}
\usepackage{slashed}
\usepackage{float}
\usepackage{setspace}
\usepackage{braket}

\usepackage{tikz}
\usetikzlibrary{decorations.markings,arrows.meta}

\DeclareMathOperator{\tr}{Tr}

\renewcommand{\Im}{\operatorname{Im}}
\newcommand{\err}[2]{\raisebox{-0.4ex}
{$\stackrel{\scriptstyle +#1}{\scriptstyle -#2}$}}

\begin{document}

\preprint{ADP26-08/T1305}

\title{Structure of QC\boldmath{${}_2$}D ground state fields at nonzero matter densities}

\author{Ragib F. Hasan}
\affiliation{Centre for the Subatomic Structure of Matter (CSSM), 
	     Department of Physics, Adelaide University, SA, 5005, Australia}

\author{Matthew Cummins}
\affiliation{Centre for the Subatomic Structure of Matter (CSSM), 
	     Department of Physics, Adelaide University, SA, 5005, Australia}

\author{Waseem Kamleh}
\affiliation{Centre for the Subatomic Structure of Matter (CSSM), 
	     Department of Physics, Adelaide University, SA, 5005, Australia}

\author{Dale Lawlor} 
\affiliation{Department of Physics, Maynooth University--National University of Ireland Maynooth, 
             County Kildare, Ireland}

\author{Derek B. Leinweber}
\affiliation{Centre for the Subatomic Structure of Matter (CSSM), 
	     Department of Physics, Adelaide University, SA, 5005, Australia}

\author{Ian van Schalkwyk}
\affiliation{Centre for the Subatomic Structure of Matter (CSSM), 
	     Department of Physics, Adelaide University, SA, 5005, Australia}

\author{Jon-Ivar Skullerud}
\affiliation{Department of Physics, Maynooth University--National University of Ireland Maynooth, 
             County Kildare, Ireland}

\begin{abstract}
A quantitative investigation into the modification of ground-state field structures in two-color
QCD (QC$_2$D) is presented at finite chemical potential. Using lattice simulations with Wilson
gauge and fermion actions, we explore the chromo-electromagnetic field strengths under varying
matter densities. To ensure accurate measurements, we develop and calibrate two highly improved
topological charge operators and evaluate four gradient flow actions.  Our results reveal a
finite-volume crossover in the regime of the anticipated phase boundary at $\mu = m_\pi/2$, with
both chromoelectric and chromomagnetic field strengths suppressed before recovering and exceeding
vacuum values at higher chemical potentials. We find the difference between the squared
chromoelectric and chromomagnetic field strengths, $E^2-B^2$, to increase in magnitude
monotonically with increasing chemical potential. At $a\mu=0.7$, we find an $11\%$ suppression of
$E^2$, a relatively small effect. A systematic analysis using sigmoid fits of lattice simulations
in the crossover regime is performed to confirm the critical chemical potential obtained from the
field structure is in agreement with the phase boundary at $m_\pi / 2$.  These findings provide new
insight into non-Abelian ground-state vacuum field structures and offer a foundation for future
studies in real QCD.
\end{abstract}

\maketitle

\section{Introduction}

The phase structure of strongly interacting matter at high densities,
beyond the nuclear saturation density and at low temperatures, remains
one of the major unresolved issues in fundamental physics, despite
extensive research efforts spanning several decades.  Gaining precise
insight into this region could help answer key questions about the
structure and behavior of compact stars, including whether deconfined
quark matter might be present within them.

Progress has been limited because conventional weak-coupling
techniques are only applicable at asymptotically high densities, and
the various theoretical models used so far lack constraints from
experimental data or first-principles calculations.  As a result,
although numerous models offer insights into potential phases and
their characteristics, no undisputed quantitative conclusions have
been reached. Lattice QCD simulations could potentially resolve many
key questions, but progress is limited by the well-known sign problem
\cite{Aarts:2015tyj}. Although progress is being made for QCD itself
(for a recent review see Ref.~\cite{Guenther:2022wcr}), simulations of
QCD-like theories that avoid this issue can offer valuable,
nonperturbative insights \cite{Itou:2025vcy}. The primary objective of
this study is to use such simulations to help constrain and inform
theoretical models.

Among the various QCD-like theories \cite{Kogut:2000ek}, two-color QCD
(QC$_2$D), which uses the SU(2) gauge group, is particularly
noteworthy
\cite{Creutz:1980zw,Hands:1999md,Braguta:2023yhd,Itou:2025vcy}. It
retains many key features of QCD, such as: confinement, dynamical
chiral symmetry breaking, and long-range interactions -- while
differing in that its baryons are bosons. Notably, the lightest baryon
in QC$_2$D is a pseudo-Goldstone boson, degenerate with the
pion. Hence, instead of a conventional nuclear matter phase, this
theory exhibits a superfluid state, marked by the condensation of
baryons, which become true Goldstone bosons at this point.

The phase structure of QC$_2$D has been the subject of numerous
investigations (see the recent reviews
\cite{Braguta:2023yhd,Itou:2025vcy} for more details), with a clear
picture emerging.

At low temperature $T$ and quark chemical potential $\mu$, we find a
hadronic phase, characterized by dynamical chiral symmetry breaking
and a vanishing Polyakov loop (indicating confinement of static color
charges).  At high $T$ and small $\mu$, there is a transition to a
quark-gluon plasma characterized by chiral symmetry restoration and a
nonvanishing Polyakov loop.  At finite, nonzero quark mass, both
these transitions are smooth crossovers, and it is not yet determined
whether their respective pseudocritical temperatures $T_\chi$ and
$T_d$ coincide at $\mu=0$.  At low $T$, as the chemical potential
increases, there is a transition to a baryon superfluid state at
$\mu=m_\pi/2$, characterized by a nonvanishing diquark condensate
$\braket{qq}$.  The superfluid transition temperature $T_{\Delta}$ is
found to be below $T_\chi$ and $T_d$,  and almost independent of
$\mu$ \cite{Boz:2013rca,Boz:2019enj}.

For $\mu$ not much above $m_\pi/2$, the superfluid phase is
characterized by a Bose-Einstein condensate of tightly bound
diquarks, described by chiral perturbation theory
\cite{Astrakhantsev:2020tdl,Iida:2024irv}.  As $\mu$ increases, there
is a crossover to a Bardeen-Cooper-Schrieffer (BCS) regime which may
be described by a Fermi sphere of weakly interacting quarks disrupted
by a diquark condensate scaling as the area of the Fermi surface,
$\braket{qq}\sim\mu^2$.  Whether or not there is a deconfinement
transition at large $\mu$ and $T=0$ remains undecided; the crossover
temperature $T_d$ is found to decrease with increasing $\mu$
\cite{Boz:2019enj,Begun:2022bxj}, but early indications of
deconfinement at large $\mu$ \cite{Hands:2006ve} were later found to
be lattice artifacts \cite{Boz:2019enj}.

A notable feature of the thermodynamics of QC$_2$D at low $T$ is that
the speed of sound $c_s$ exhibits a peak in the superfluid region
where it exceeds its high-density (conformal) limit, $c_s^2=1/3$
\cite{Iida:2022hyy}.  This feature is also found in isospin-QCD
\cite{Abbott:2024vhj}, and is inferred from neutron star observations
to also be present in real QCD \cite{Annala:2019puf,Brandes:2022nxa}
(see also \cite{Fukushima:2024gmp} for a discussion of the relation
between these theories).

Of particular interest is the gluodynamics and ground-state field
structures of QC$_2$D, which may provide direct lessons for real QCD.
Like QCD, QC$_2$D is a non-Abelian gauge field theory enabling gluons
to self-interact. This makes the empty vacuum unstable to the
formation of quark and gluon condensates which permeate spacetime.
These ground-state fields form the foundation of matter and give rise
to quark confinement and dynamical chiral symmetry breaking
\cite{Gross:2022hyw}.

Although transitioning from QCD to QC$_2$D fundamentally changes the
structure of hadrons, where baryons in QC$_2$D consist of two valence
quarks, the gluonic sector is at most indirectly affected. Studies
quantitatively comparing gluon and ghost propagators in SU(2) and
SU(3) gauge theory report quantitative agreement between the
propagators over a wide range of momenta
\cite{Cucchieri:2007ji,Sternbeck:2007ug}, confirming an anticipated
infrared independence from the gauge group \cite{Alkofer:2000wg}.
This stability allows researchers to explore gluonic observables such
as the gluon propagator and string tension
\cite{Hands:2006ve,Boz:2013rca,Boz:2018crd,Begun:2022bxj} within
QC$_2$D, potentially offering valuable insights into their behavior in
real QCD. Consequently, studying quantities like the energy density,
topological charge density, and the chromoelectric and chromomagnetic
field strength densities in QC$_2$D are of interest.

Topological charge and susceptibility has previously been studied in
dense QC$_2$D by several groups
\cite{Alles:2006ea,Hands:2011hd,Astrakhantsev:2019wnp,Astrakhantsev:2020tdl,Iida:2019rah}
(see Ref.~\cite{Lombardo:2021jpn} for an overview of topological
aspects of dense matter).  For $N_f=2$ at low temperature, simulations
with Wilson fermions find no evidence that the topological
susceptibility changes with chemical potential
\cite{Hands:2011hd,Iida:2019rah}, while simulations with staggered
fermions at somewhat higher temperature find a clear drop for
$\mu\gtrsim m_\pi/2$ \cite{Astrakhantsev:2020tdl}.  A study of the
chromoelectric flux tube \cite{Ishiguro:2021yxr} found that the flux
tube persists in the superfluid phase.

This paper complements and extends these studies by exploring how the
Euclidean action density, the topological charge density, and the
squared chromo-electric and -magnetic field strengths $\braket{E^2}$
and $\braket{B^2}$ respectively change under the influence of a
nonzero matter density.  We will focus in particular on $\braket{E^2}$
and $\braket{B^2}$ (which were also studied in
Ref.~\cite{Astrakhantsev:2020tdl}) since we find that these quantities
are sensitive to the onset transition at low $T$.

As these gluonic measures are calculated directly from the vacuum
gauge fields, one needs to smooth the gauge fields to suppress
otherwise large and uncontrolled renormalizations.
Gradient flow is now the accepted standard for gauge field smoothing
and this is used herein.  However, a number of different
implementations have been proposed, drawing on a variety of
improvement schemes corresponding to different actions.  Four of the
most popular schemes are carefully explored herein.  The techniques
used to select an optimal smoothing are reviewed before presenting our
best understanding of how the finite matter density induces changes in
the QC$_2$D ground-state field configurations.

The outline of this paper is as follows.  Section
\ref{sec:gaugeFields} briefly reviews and summarizes the lattice
techniques used to implement a finite matter density in QC$_2$D.  We
then proceed with a qualitative look at the vacuum field structure and
the manner in which a finite matter density changes that structure in
Sec.~\ref{sec:visualisations}. Moving toward a quantitative
examination, Sec.~\ref{sec:gradientFlowAction} examines the features
of various lattice-action improvement schemes, determining a stable
action that preserves nonperturbative field structure and quantifying
the minimal gradient flow required to extract quantitative results
from the gauge-field configurations.  Our survey of the impact of a
finite matter density on the ground-state field structure of QC$_2$D
are presented in Sec.~\ref{sec:survey}.  Here we examine the critical
chemical potential of the finite-density transition through an
inflection-point analysis drawing on additional simulations near the
finite-volume crossover. We find the critical chemical potential
obtained from changes in the ground-state field structure to coincide
with the phase boundary at $m_\pi/2$.  A summary of our best
understanding of the impact of finite density on the ground-state
fields is presented in the conclusions of Sec.~\ref{sec:conclusions}.

\section{Finite density in QC$_2$D}
\label{sec:gaugeFields}

This study employs the conventional Wilson plaquette gauge action alongside two flavors of Wilson
fermions to investigate QC$_2$D. The fermion sector includes a gauge- and isospin-singlet diquark
source term, $j$, which serves to lift the low-lying eigenvalues of the Dirac operator. This
modification enables a controlled exploration of diquark condensation phenomena
\cite{Cotter:2012mb}. The corresponding quark action is
\begin{equation}
S_Q+S_J=\sum_{i=1,2}\bar\psi_iM\psi_i
 + \kappa j[\psi_2^{\rm tr}\, (C\gamma_5)\,\tau_2\,\psi_1-{\rm H.c.}],
\label{eq:Slatt}
\end{equation}
where $\rm tr$ denotes the transpose, $\kappa = 1/(2 m_q + 8r)$ is the hopping parameter for quark
mass $m_q$ and Wilson parameter $r=1$. The matrix $C$ is the charge conjugation matrix, $\tau_2$ is
the second Pauli matrix acting in color space,
and the fermion matrix $M$ is
\begin{align}
M_{xy}=\delta_{xy}-\kappa\sum_\nu&\Bigl[(1-\gamma_\nu)\,e^{\mu\delta_{\nu0}}\,U_\nu(x)\,\delta_{y,x+\hat\nu}\nonumber\\ 
&+(1+\gamma_\nu)\,e^{-\mu\delta_{\nu0}}\,U^\dagger_\nu(y)\,\delta_{y,x-\hat\nu}\Bigr]\, ,
\label{eq:Mwils}
\end{align}
where $\gamma_\nu$ are the Hermitian gamma matrices.
We note here the special role the time component plays in implementing the chemical potential.

To understand the origin of this fermion action, we recall the continuum formulation for
noninteracting fermions at temperature $T$
\begin{align}
S = \int_0^{1/T}d\tau \int d^3x\, \bar{\psi}(\gamma_\nu \partial_\nu + m)\psi\, .
\end{align}
To obtain the grand canonical partition function, we add the chemical potential term with number
operator $N$ as
\begin{align}
\frac{\mu N}{T} 
&= \int_0^{1/T}d\tau \int d^3 x\, \mu\, \bar{\psi} \gamma_4 \psi \, .
\end{align}
Incorporating a gauge field $A_\nu$, the action becomes
\begin{align}
S = \int_0^{1/T} d\tau \int d^3x\, \bar{\psi}[\gamma_\nu (\partial_\nu + i A_\nu) + \mu \gamma_4 + m]\psi\, .
\label{eq:partAction}
\end{align}
Reference~\cite{Aarts:2015tyj} highlights several key features of this expression. Notably, $\mu$ enters
similarly to $iA_4$, the imaginary-time component of a four-vector field. This suggests that
finite-density effects may manifest in the temporal direction, potentially revealing differences
between chromoelectric and chromomagnetic fields. 

Additionally, at $\mu = 0$, the Dirac operator satisfies $\gamma_5$-Hermiticity
\begin{align}
(\gamma_5 M)^\dagger = \gamma_5 M \implies M^\dagger = \gamma_5 M \gamma_5\, ,
\end{align}
which implies
\begin{align}
\det(M^\dagger) = \det(\gamma_5 M \gamma_5) = \det M =(\det M)^*
\end{align}
and hence the determinant is real. 
However, for $\mu \neq 0$,
\begin{equation}
M^\dagger (\mu) = \gamma_5 M(-\mu^*)\gamma_5 \, ,
\end{equation}
and
\begin{equation}\label{eq:MM}
(\det M(\mu))^* = \det M(-\mu^*) \in \mathbb{C} \, .
\end{equation}
In general, this can lead to a complex probability weight known as the sign problem, obstructing
importance sampling.

In two-color QC$_2$D \cite{Murakami:2023ejc}, the absence of the sign problem comes from the
special property that the fundamental representation of SU(2) is pseudoreal
\cite{Kogut:2000ek,Hands:2001jn}. This means that quarks and antiquarks are related in a way that
ensures the Dirac operator, even with a finite chemical potential, is always mapped to its complex
conjugate by a similarity transformation. As a result, the eigenvalues of the Dirac operator occur
in complex-conjugate pairs, which ensures the fermion determinant remains real. For an even number
of quark flavors, the determinant is also non-negative, so the path integral measure can still be
interpreted probabilistically. 

A na\"ive attempt to add $\mu \, \bar{\psi}\gamma_4\psi$ directly to the lattice action results in
ultraviolet divergences. Instead, inspired by the continuum analogy, one treats $\mu$ as the
imaginary component of an Abelian vector field in the temporal direction. This leads to the
introduction of the chemical potential in the temporal hopping terms
\cite{Hasenfratz:1983ba,Kogut:1983ia,Aarts:2015tyj} of Eq.~(\ref{eq:Mwils}), ensuring that
forward-moving quarks in time are favored, while backward-moving ones are suppressed, creating an
asymmetry in temporal propagation.

The chemical potential represents the free energy required to
introduce a particle into the system. Consequently, in QC$_2$D,
observable changes are expected when $\mu$ reaches half of the diquark
baryon mass, $m_b$. As $\mu$ is associated with a single quark and
baryons have two quarks (in QC$_2$D), $\mu_C = m_b/2 = m_\pi/2$ as
$m_b = m_\pi$ in QC$_2$D. Due to this degeneracy with the pion mass,
$m_\pi/2$ marks the onset of particle creation from the vacuum and
potential structural changes in the ground-state vacuum fields.

In this work we will be using the ``Coarse'' ($\beta=1.9,\, \kappa=0.1680$) and ``Fine''
($\beta=2.1,\, \kappa=0.1577$) lattice ensembles described in Ref.~\cite{Boz:2019enj}.  We use the
lowest-temperature lattices for each ensemble, with lattice volumes of $16^3\times24$ and
$16^3\times32$ respectively, both corresponding to temperatures $T\approx40\,$MeV.  The lattice
spacings, set by the string tension of $\sqrt{\sigma}=440\,\mathrm{MeV}$, are $a=0.178(6)\,$fm and
$a=0.130(1)\,$fm \cite{Hands:2024nkx} respectively, with pion masses $am_\pi=0.645(8)$ and
$am_\pi=0.446(3)$ respectively, both corresponding to a pion mass $m_\pi\approx700\,$MeV.

\section{First look at finite density field modification}
\label{sec:visualisations}

We begin by visualizing the action density, topological charge density and distributions of the
squared chromo-electric and -magnetic fields strengths, $E^2(x)$ and $B^2(x)$ respectively, on
representative gauge field configurations.  The five-loop ${\cal O}(a^4)$-improved lattice
field-strength tensor \cite{Bilson-Thompson:2002xlt} is used to determine these from the lattice
gauge fields.

We consider the vacuum field structure on the $\beta=1.9,\, \kappa=0.1680$
lattices at $a \mu = 0$ and contrast this with a representative
$a\mu=0.4$ configuration above the phase boundary at
$am_\pi/2=0.323(4)$.  This is then compared with a representative
$a\mu=0.7$ configuration, where lattice artifacts associated with
large $a\mu$ are expected to be benign.  We seek to observe
qualitative changes in the vacuum field structure.

To suppress short distance fluctuations in exposing the longer distance nonperturbative structures
in the gauge fields, we implement a standard gradient flow and illustrate the field structures at a
moderate gradient flow time of $t_g = 2.4$.
To see into the 3D volume, a threshold is set such that field values below the threshold are not
rendered.  In all cases of comparison, the threshold is held fixed.

As the Euclidean time dimension is unique at finite density, it is important to include this
dimension in the visualizations.  Accordingly, we slice the four-dimensional volume at constant
values of the $x$ coordinate to render a 3D perspective. Thus the visualizations illustrate
the $y$, $z$ and $t$ variation of the field structure.  The time direction is easily identified as
the elongated axis with 24 units with respect to the other axes of length 16.  While many
configurations and many different $x$ coordinates were examined, the visualizations presented here
are selected as good representatives of our general observations.

\begin{figure*}[t]
\centering
\includegraphics[width=0.32\textwidth]{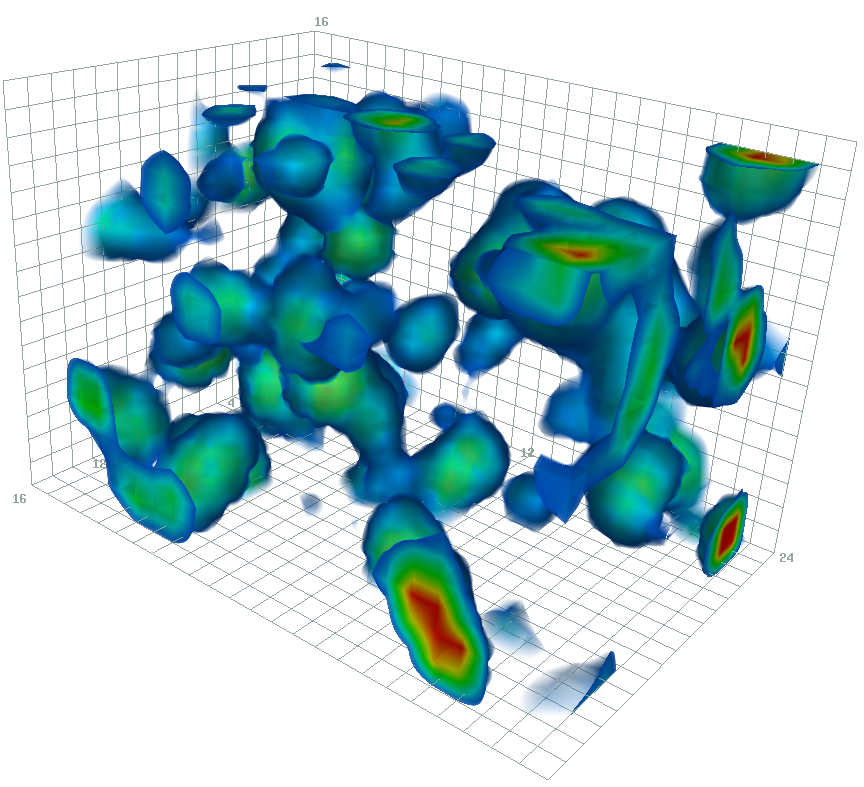} \qquad
\includegraphics[width=0.32\textwidth]{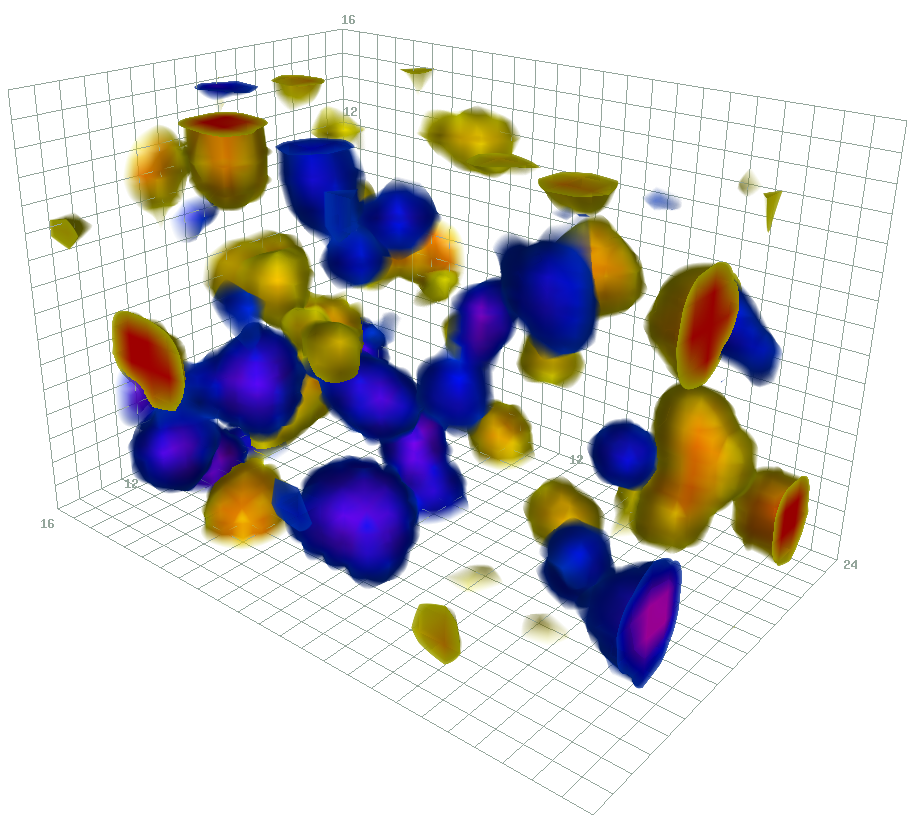} \qquad
\includegraphics[width=0.32\textwidth]{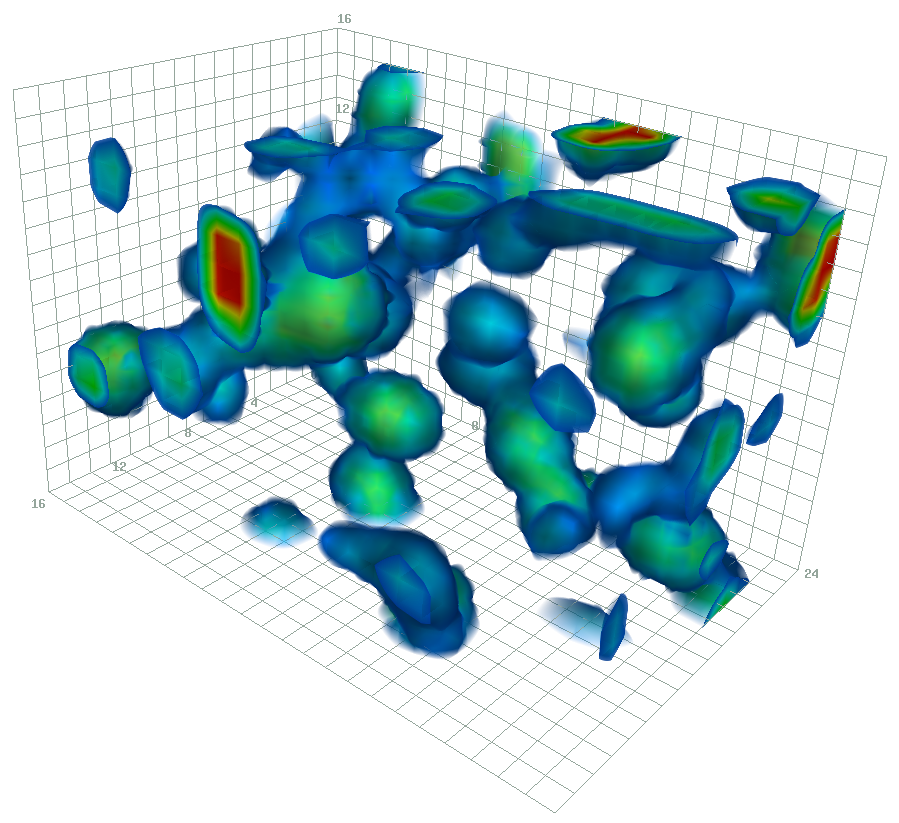} \qquad
\includegraphics[width=0.32\textwidth]{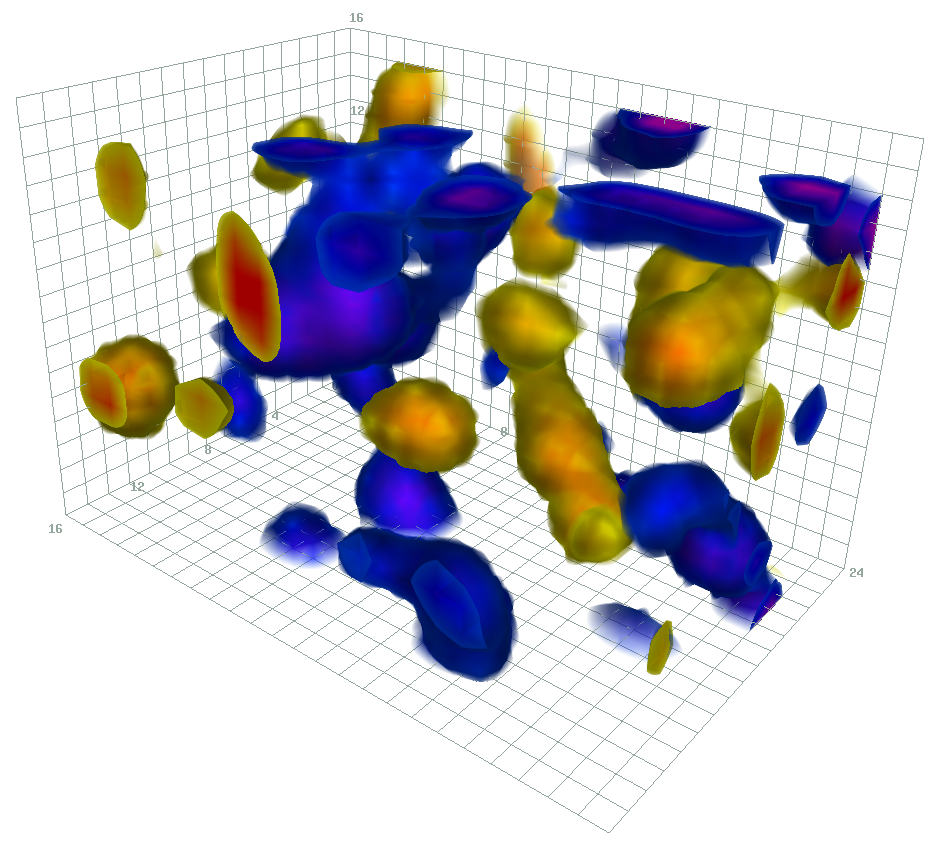} \qquad
\includegraphics[width=0.32\textwidth]{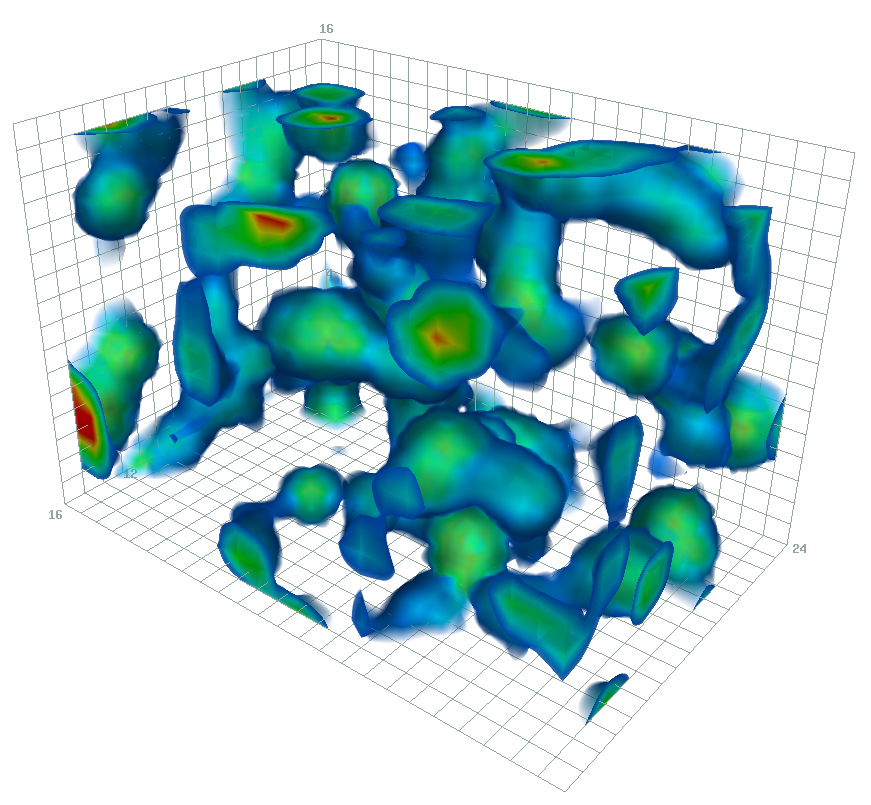} \qquad
\includegraphics[width=0.32\textwidth]{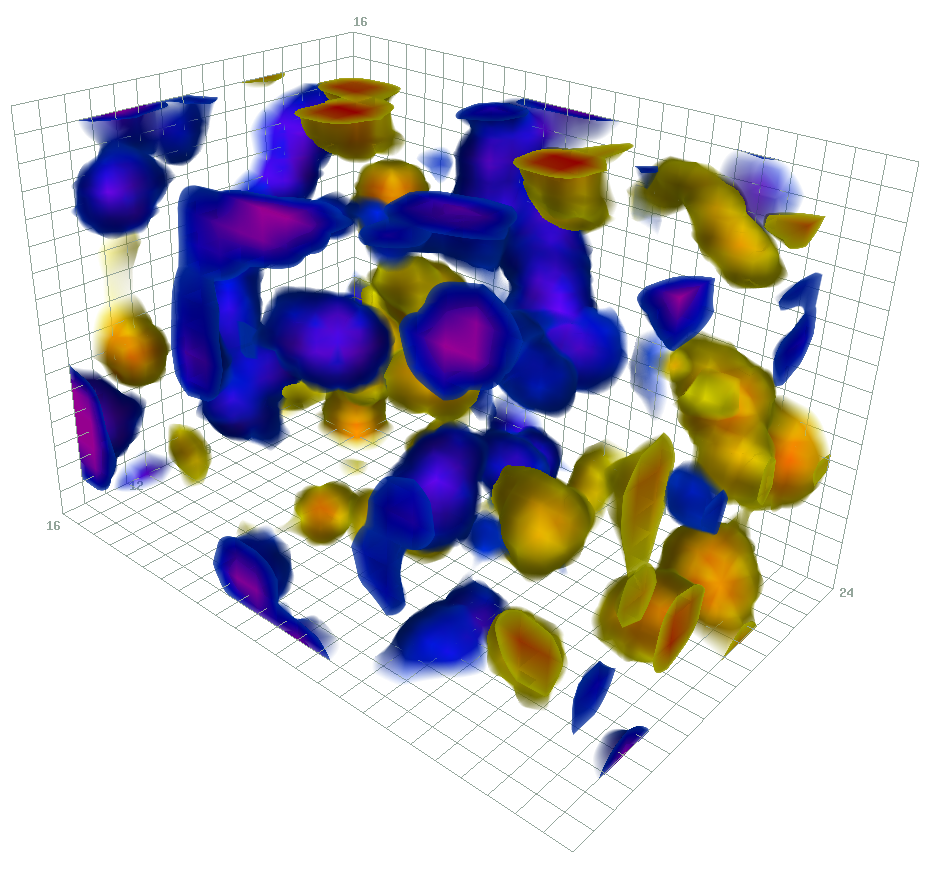} \qquad
\caption{Comparison of the action density (left column) and topological charge density (right
  column) illustrated for chemical potentials of $a \mu = 0$ (top), $a \mu = 0.4$ (middle) and $a
  \mu = 0.7$ (bottom).  For the action density in the left column, red/yellow/green/blue shading
  indicates high to low local action density. Similarly, for the topological charge density in the
  right column, violet/blue indicates negative topological charge density and red/yellow indicates
  positive topological charge density, with violet and red illustrating the largest magnitudes.
  The lowest action and topological-charge density regions below their respective thresholds are
  not rendered as described in the text.}
\label{fig:actionCharge}
\end{figure*}

The left column of Fig.~\ref{fig:actionCharge} shows the action densities of three configurations
at the aforementioned three chemical potentials. The corresponding topological charge densities are
illustrated in the right column.  Despite very large differences in the chemical potentials
considered in generating these configurations, the renderings are qualitatively similar, confirming
early expectations that the vacuum structure modification is subtle
\cite{Hands:2011hd,Boz:2018crd}.  In a blind test, it would be very difficult to tell these
configurations apart.
Studies of the topological susceptibility also indicate an insensitivity to the chemical potential
\cite{Iida:2019rah,Iida:2024irv}.

\begin{figure*}[t]
\centering
\includegraphics[width=0.32\textwidth]{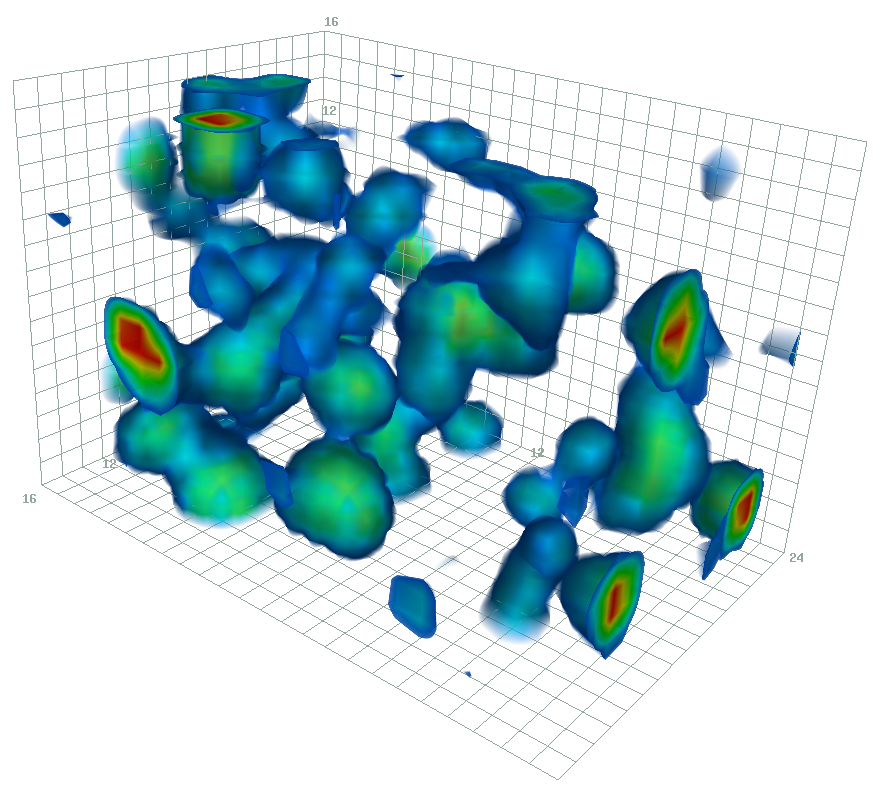}
\includegraphics[width=0.32\textwidth]{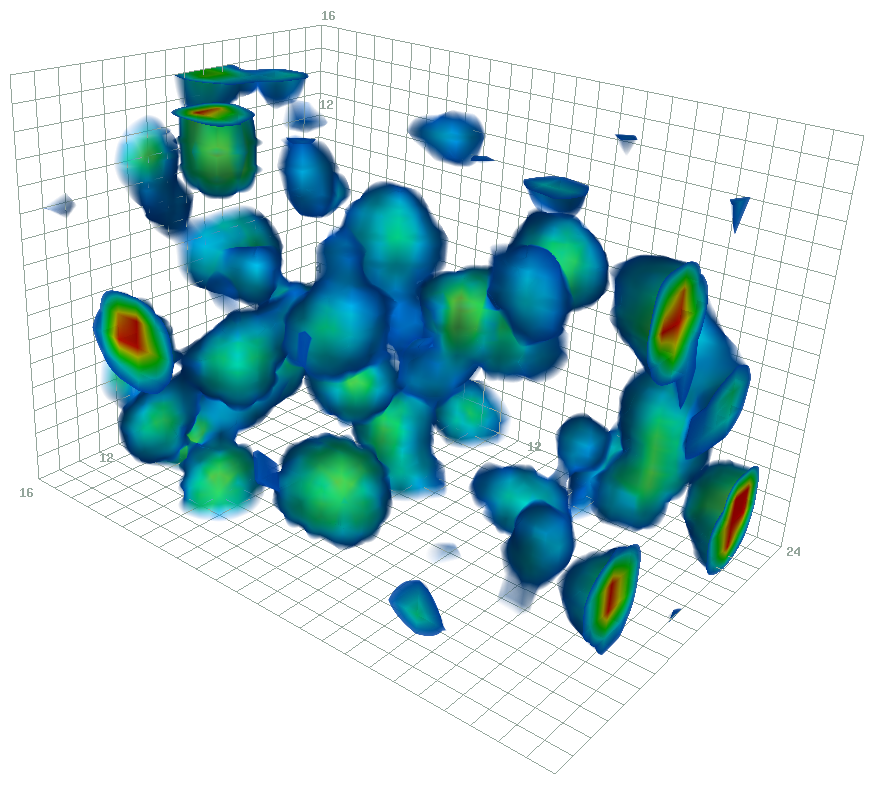}
\includegraphics[width=0.32\textwidth]{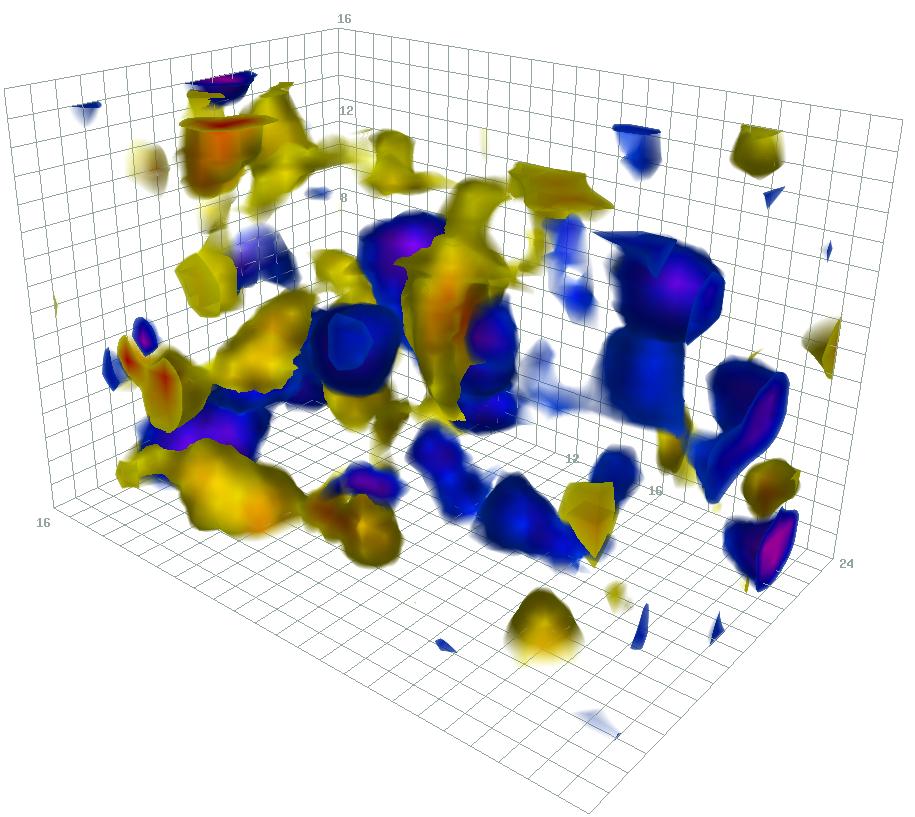}
\includegraphics[width=0.32\textwidth]{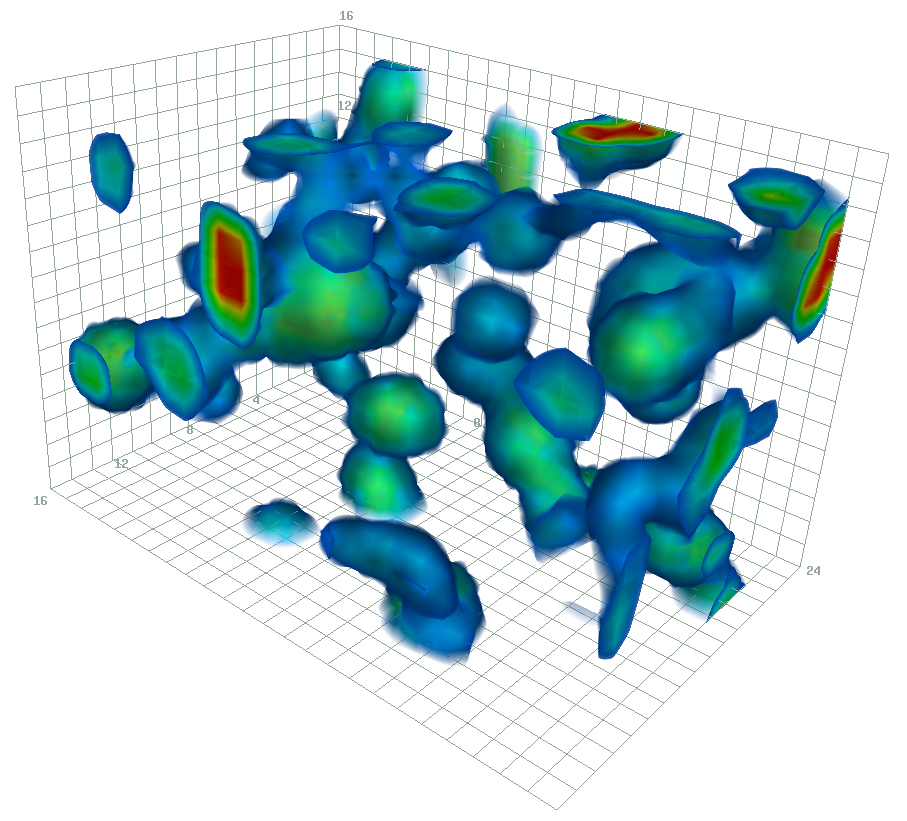}
\includegraphics[width=0.32\textwidth]{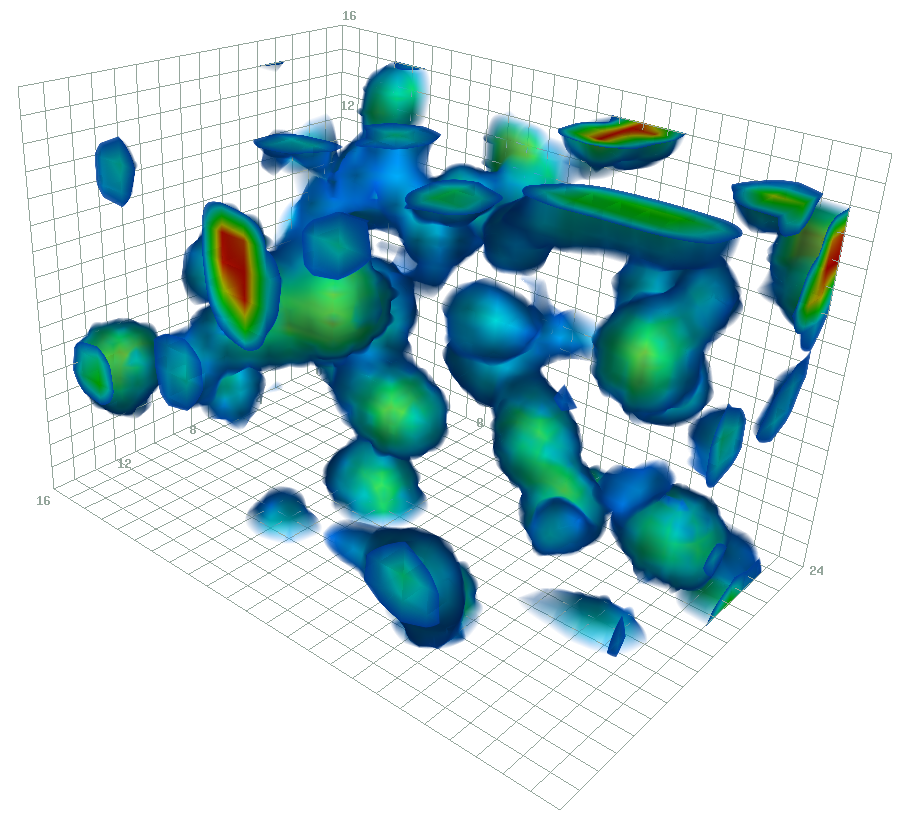}
\includegraphics[width=0.32\textwidth]{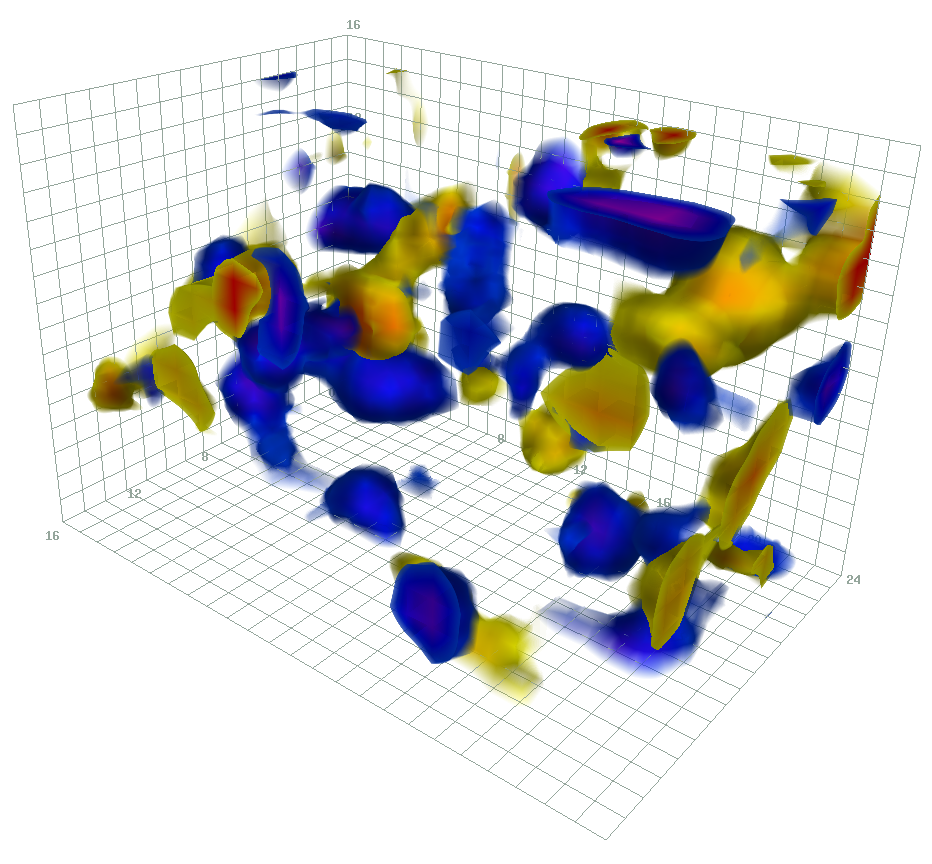}
\includegraphics[width=0.32\textwidth]{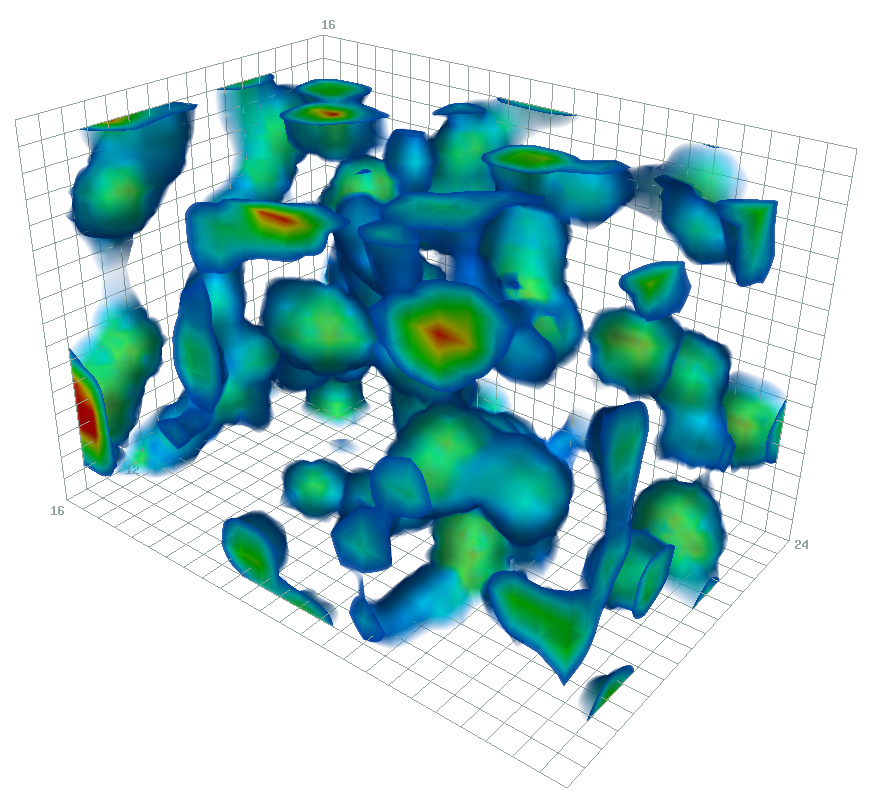}
\includegraphics[width=0.32\textwidth]{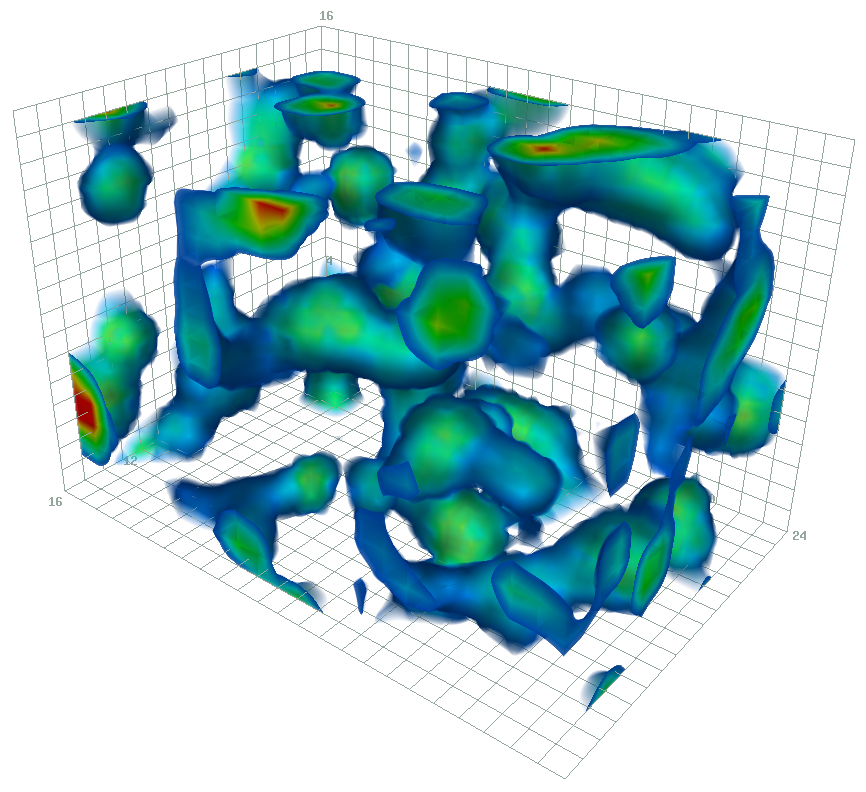} 
\includegraphics[width=0.32\textwidth]{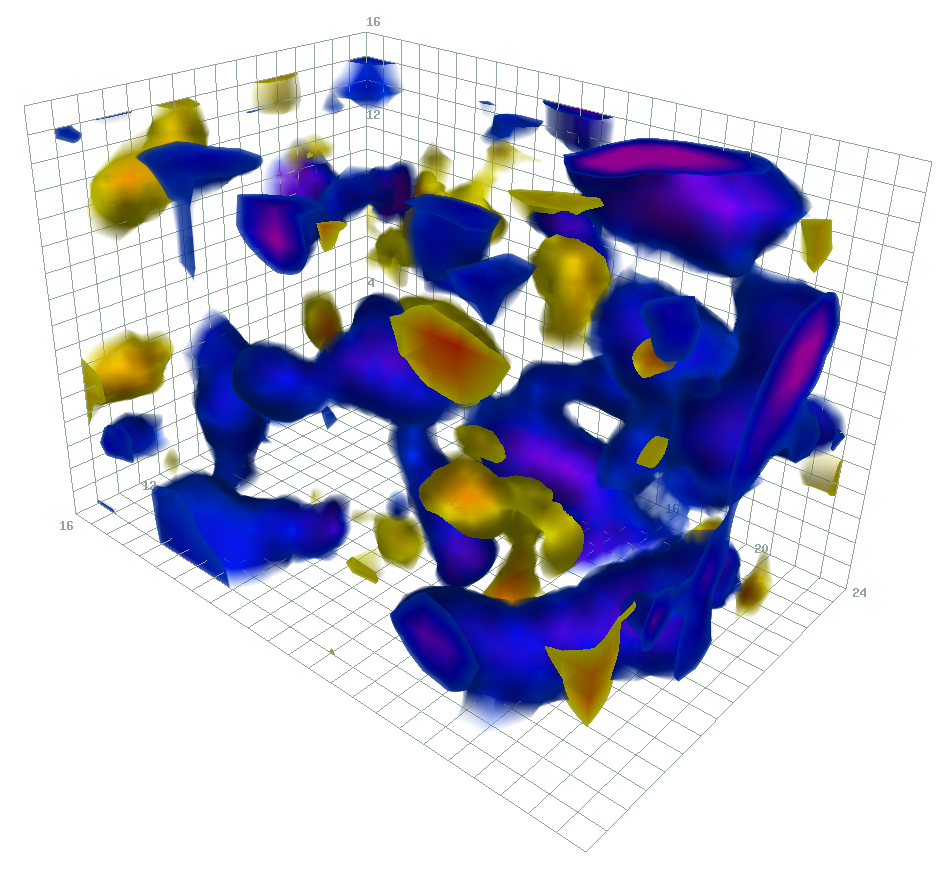} 
\caption{Comparison of squared chromoelectric field, $E^2(x)$ (left column), the squared
  chromomagnetic field, $B^2(x)$ (middle column), and the difference, $E^2(x) - B^2(x)$ (right
  column) for chemical potentials of $a \mu = 0$ (top row), $a \mu = 0.4$ (middle row) and $a \mu =
  0.7$ (bottom row). Color shadings for the positive squared field strengths, $E^2(x)$ and
  $B^2(x)$, are as for the action density of Fig.~\ref{fig:actionCharge}. The difference $E^2(x) -
  B^2(x)$ has both positive and negative values and is rendered in the same manner used for the
  topological charge in Fig.~\ref{fig:actionCharge}. Thus, violet/blue indicates a negative value
  [i.e. $B^2(x) > E^2(x)$] and red/yellow indicates a positive value [i.e. $E^2(x) > B^2(x)$]. An
  abundance of violet/blue in the bottom right plots reveals an impact of the chemical potential on
  the balance of $E^2(x)$ and $B^2(x)$.}
\label{fig:EandB}
\end{figure*}

Figure \ref{fig:EandB} illustrates the squared field strengths, $E^2(x)$ and $B^2(x)$, for the same
three configuration slices. It is difficult by direct comparison to tell these fields apart with
both maintaining approximately the same shape of objects. To more clearly see the difference in
$E^2(x)$ and $B^2(x)$, we visualize $E^2(x) - B^2(x)$ in the right-most column of
Fig.~\ref{fig:EandB}. Here we finally observe a qualitative impact of the chemical potential on the
ground-state field structure.  The large chemical potential configurations show an abundance of
violet/blue revealing a qualitative-level impact of the chemical potential on the balance of
$E^2(x)$ and $B^2(x)$.

Thus, the nonzero chemical potential has indeed impacted the vacuum balance of
chromo-electromagnetic field strengths.  This signature demands a quantitative analysis to learn
the nature of the dynamics.  In proceeding to do this, we will examine the utility of several
improved gauge actions under consideration for the gradient flow.  We will also construct
alternative improved operators for the topological charge density and use this as a measure of
systematic errors in the determinations of the chromo-electromagnetic field strengths.  With these
improvements in hand we will determine the minimal gradient flow required to determine the fields
and discover the impact of the chemical potential on the ground-state fields.

\section{Lattice-action improvement in the gradient flow}
\label{sec:gradientFlowAction}

\subsection{Highly improved operators for measuring ground-state field strengths}
\label{subsec:impOps}

\tikzset{->-/.style={decoration={
  markings,
  mark=at position .68 with {\arrow{Latex}}},postaction={decorate}}}

\tikzset{-->--/.style={decoration={
  markings,
  mark=at position .59 with {\arrow{Latex}}},postaction={decorate}}}

\def\mbynclover{\tikz[baseline=-0.6ex]{
\fill (0.5ex,0.5ex) circle (1.5pt) coordinate (A);
\fill (8.5ex,0.5ex) coordinate (B);
\fill (8.5ex,4.5ex) coordinate (C);
\fill (0.5ex,4.5ex) coordinate (D);
\draw[black,-->--] (A)--(B);
\draw[black,->-] (B)--(C);
\draw[black,-->--] (C)--(D);
\draw[black,->-] (D)--(A);

\fill (-0.5ex,0.5ex) circle (1.5pt) coordinate (A);
\fill (-0.5ex,4.5ex) coordinate (B);
\fill (-8.5ex,4.5ex) coordinate (C);
\fill (-8.5ex,0.5ex) coordinate (D);
\draw[black,->-] (A)--(B);
\draw[black,-->--] (B)--(C);
\draw[black,->-] (C)--(D);
\draw[black,-->--] (D)--(A);

\fill (0.5ex,-0.5ex) circle (1.5pt) coordinate (A);
\fill (0.5ex,-4.5ex) coordinate (B);
\fill (8.5ex,-4.5ex) coordinate (C);
\fill (8.5ex,-0.5ex) coordinate (D);
\draw[black,->-] (A)--(B);
\draw[black,-->--] (B)--(C);
\draw[black,->-] (C)--(D);
\draw[black,-->--] (D)--(A);

\fill (-0.5ex,-0.5ex) circle (1.5pt) coordinate (A);
\fill (-8.5ex,-0.5ex) coordinate (B);
\fill (-8.5ex,-4.5ex) coordinate (C);
\fill (-0.5ex,-4.5ex) coordinate (D);
\draw[black,-->--] (A)--(B);
\draw[black,->-] (B)--(C);
\draw[black,-->--] (C)--(D);
\draw[black,->-] (D)--(A);}
}

\def\nbymclover{\tikz[baseline=-0.6ex]{
\fill (0.5ex,0.5ex) circle (1.5pt) coordinate (A);
\fill (4.5ex,0.5ex) coordinate (B);
\fill (4.5ex,8.5ex) coordinate (C);
\fill (0.5ex,8.5ex) coordinate (D);
\draw[black,->-] (A)--(B);
\draw[black,-->--] (B)--(C);
\draw[black,->-] (C)--(D);
\draw[black,-->--] (D)--(A);

\fill (-0.5ex,0.5ex) circle (1.5pt) coordinate (A);
\fill (-0.5ex,8.5ex) coordinate (B);
\fill (-4.5ex,8.5ex) coordinate (C);
\fill (-4.5ex,0.5ex) coordinate (D);
\draw[black,-->--] (A)--(B);
\draw[black,->-] (B)--(C);
\draw[black,-->--] (C)--(D);
\draw[black,->-] (D)--(A);

\fill (0.5ex,-0.5ex) circle (1.5pt) coordinate (A);
\fill (0.5ex,-8.5ex) coordinate (B);
\fill (4.5ex,-8.5ex) coordinate (C);
\fill (4.5ex,-0.5ex) coordinate (D);
\draw[black,-->--] (A)--(B);
\draw[black,->-] (B)--(C);
\draw[black,-->--] (C)--(D);
\draw[black,->-] (D)--(A);

\fill (-0.5ex,-0.5ex) circle (1.5pt) coordinate (A);
\fill (-4.5ex,-0.5ex) coordinate (B);
\fill (-4.5ex,-8.5ex) coordinate (C);
\fill (-0.5ex,-8.5ex) coordinate (D);
\draw[black,->-] (A)--(B);
\draw[black,-->--] (B)--(C);
\draw[black,->-] (C)--(D);
\draw[black,-->--] (D)--(A);}
}

On the original Monte Carlo generated configurations, the lattice field strength tensor operators
encounter renormalization factors far from unity. The application of gradient flow significantly
suppresses the ultraviolet fluctuations and brings the renormalization factors toward 1. This
process rapidly alters the lattice field strengths within just a few updates. As a result,
determining the optimal flow time for analyzing the field strengths becomes a nuanced challenge.

The results of the previous Section place our focus on the squared chromo-electromagnetic field
strengths, $E^2(x)$ and $B^2(x)$, and their differences. To determine the gauge field smoothness
required to accurately determine these field strengths, we focus on the topological charge density,
$q(x)$. As $E^2(x)$, $B^2(x)$ and $q(x) \propto \vec E(x) \cdot \vec B(x)$ are all 
constructed from the same flowed lattice field strength tensor, the topological charge density
provides a single operator that is sensitive to both the chromo-electric and -magnetic field flows.

To determine the flow time where discretization artifacts in the flowed field strength tensor are
sufficiently small, we follow the ideas introduced in Ref.~\cite{Mickley:2023exg} and consider two
highly-improved measures of the topological charge density. One approach improves the lattice field
strength tensor first \cite{Bilson-Thompson:2002xlt} and then derives the topological charge
density from this refined operator; we call this estimate the ``Improved $F_{\mu\nu}$" density,
$q_F(x)$. The alternative method directly improves the topological charge density itself
\cite{deForcrand:1997esx}, referred to as the ``Direct" density, $q_D(x)$. Although both techniques
yield ${\cal O}(a^4)$-improved estimates of the topological charge density, they differ
significantly in their ${\cal O}(a^6)$ error structures. Our objective is to identify the regime
where these discrepancies become negligible, enabling controlled measurements of the
chromoelectric and chromomagnetic field strengths.

Here we briefly summarize the construction of these two distinct $\mathcal{O}(a^4)$-improved
topological charge operators, using the conventions presented in Ref.~\cite{Mickley:2023exg}.
The first operator, termed the ``Improved $F_{\mu\nu}$'' scheme [$q_F(x)$], is constructed by
inserting an $\mathcal{O}(a^4)$-improved field strength tensor into the topological charge
definition
\begin{equation}
    q(x) = \frac{g^2}{32\pi^2} \epsilon_{\mu\nu\rho\sigma} \tr\big(F_{\mu\nu}(x) F_{\rho\sigma}(x)
    \big) \,. 
\end{equation}
The $m\times n$ clover term $\mathcal{C}_{\mu\nu}^{(m\times n)}(x)$ is defined as the sum of
$m\times n$ and $n\times m$ Wilson loops in the $\mu$-$\nu$ plane at the point $x$
\begin{equation}
\label{eq:clover}
    \mathcal{C}_{\mu\nu}^{(m\times n)} = \;\mbynclover \;\,+\,\; \nbymclover \;\;.
\end{equation}
Each clover term estimates the field strength tensor as
\begin{equation} \label{eq:cloverfmunu}
    F_{\mu\nu}^{(m\times n)}(x) = \frac{1}{8} \Im\big(\mathcal{C}_{\mu\nu}^{(m\times n)}(x) \big) \,.
\end{equation}
The improved operator is then formed by a linear combination
\begin{equation}
    ga^2F_{\mu\nu}(x) = \sum_{m,n} k^{(m\times n)} F_{\mu\nu}^{(m\times n)}(x) \,.
\end{equation}
To cancel $\mathcal{O}(a^2)$ and $\mathcal{O}(a^4)$ errors, $(m,n)=(1,1)$, $(2,2)$, $(1,2)$,
$(1,3)$, and $(3,3)$ are considered with coefficients
\begin{equation}
\begin{aligned}
    k^{(1\times 1)} &= \frac{19}{9} - 55\,k^{(3\times 3)} \,, \\
    k^{(2\times 2)} &= \frac{1}{36} - 16\,k^{(3\times 3)} \,, \\
    k^{(1\times 2)} &= -\frac{32}{45} + 64\,k^{(3\times 3)} \,, \\
    k^{(1\times 3)} &= \frac{1}{15} - 6\,k^{(3\times 3)} \,,
\end{aligned}
\end{equation}
with $k^{(3\times 3)}$ as a free parameter \cite{Bilson-Thompson:2002xlt}.

Alternatively, one can define a set of discretized topological charge operators $q^{(m\times
  n)}(x)$ for each clover term \cite{deForcrand:1997esx}
\begin{equation}
    q^{(m\times n)} = \frac{1}{32\pi^2} \frac{1}{m^2 n^2} \varepsilon_{\mu\nu\rho\sigma} \tr\big(
    F_{\mu\nu}^{(m\times n)} F_{\rho\sigma}^{(m\times n)} \big) \,. 
\end{equation}
Here, $F_{\mu\nu}^{(m\times n)}(x)$ is as before. These can be combined into another improved
operator 
\begin{equation}
    q_D(x) = \sum_{m,n} c^{(m\times n)} q^{(m\times n)}(x) \,.
\end{equation}
This is the ``Direct" improvement scheme ($q_D(x)$). Since $q(x)$ is nonlinear in $F_{\mu\nu}$, the
two schemes will have different ${\cal O}(a^6)$ errors. Using the same five clover terms, the
coefficients which cancel ${\cal O}(a^2)$ and ${\cal O}(a^4)$ errors are
\begin{equation}
\begin{aligned}
    c^{(1\times 1)} &= \frac{1}{9}\big(19 - 55\,c^{(3\times 3)}\big) \,, \\
    c^{(2\times 2)} &= \frac{1}{9}\big(1 - 64\,c^{(3\times 3)}\big) \,, \\
    c^{(1\times 2)} &= \frac{1}{45}\big(\!-\!64 + 640\,c^{(3\times 3)}\big) \,, \\
    c^{(1\times 3)} &= \frac{1}{5} - 2\,c^{(3\times 3)} \,.
\end{aligned}
\end{equation}
These match the coefficients used in constructing an improved action from the same loops
\cite{deForcrand:1997esx,deForcrand:1995qq}.

To compare the schemes, we examine three-loop and five-loop versions of both. The three-loop
versions use $k^{(3\times 3)}=1/90$ and $c^{(3\times 3)}=1/10$, such that the $1 \times 2$ and
$1\times 3$ loops do not contribute. The five-loop versions use $k^{(3\times 3)}=1/180$
\cite{Bilson-Thompson:2002xlt} and $c^{(3\times 3)}=1/20$
\cite{deForcrand:1997esx,deForcrand:1995qq}.

We have compared the three-loop and five-loop schemes for both $q_F(x)$ and $q_D(x)$. For each, we
compute the total sum of the absolute topological charge density $Q=\sum |q(x)|$ and the relative
error
\begin{equation}
    R_E = \frac{|Q_F - Q_D|}{\frac{1}{2}(Q_F + Q_D)} \,.
\end{equation}
Our calculations show that the five-loop improvement schemes are more consistent than the
three-loop schemes. Thus, we utilize the five-loop versions of both $q_F(x)$ and $q_D(x)$ as our
benchmark for identifying when optimal smoothing is achieved.

\subsection{Gradient flow}
\label{subsec:gf}

The Yang-Mills gradient flow \cite{Luscher:2010iy} evolves the gauge field $A_\mu(x)$ into a flowed
field $A_\mu(x,t)$ via a diffusion equation in flow time $t$ 
\begin{equation}
\frac{d}{dt}\, A_\mu(x;t) = -\frac{\delta S}{\delta A_\mu^a(x;t)}\, t^a 
                          = D_\nu\, F_{\nu\mu}(x;t) \, ,
\label{eq:gflow_continuum}
\end{equation}
with
\begin{equation}
\qquad A_\mu(x;t) = A_\mu^a(x;t)\,t^a \, , \mbox{ and } A_\mu(x;t=0) = A_\mu(x)\, .
\label{eq:gflow_defs}
\end{equation}
The $t^a$ are the generators of SU($N$) gauge theory,
$F_{\mu\nu} = \partial_\mu A_\nu - \partial_\nu A_\mu + [A_\mu,A_\nu]$ is the field strength
of the flowed field, and 
$D_\mu = \partial_\mu + [A_\mu,\cdot\,]$ is the flow-time covariant derivative, the dot
denoting the argument on which $D_\mu$ acts. 

The smeared gauge field is the solution to this differential equation for $t>0$.  The links evolve
in the direction of steepest descent in the action and the action is monotonically decreasing
in flow time.
The flow is gauge covariant and its diffusive character smooths the gauge field over
a physical radius $d=\sqrt{8t}$, exponentially damping momentum modes above the scale
$\mu = 1/d$.

On the lattice we work with the dimensionless flow time $t_g \equiv t/a^2$, such that
the physical flow time is $t = t_g\, a^2$ and the physical smoothing radius is
$d = \sqrt{8t} = \sqrt{8 t_g\, a^2}$.
The flow is implemented with reference to the flowed link variables $U_\mu(x;t)$,
and their associated gauge action $S[U]$.  Equation~(\ref{eq:gflow_continuum}) takes the form
\begin{equation}
\frac{d}{dt_g} U_\mu(x;t_g) = -ig^2\, t^a\, \nabla^a_{x,\mu}\, S[U]\, U_\mu(x;t_g) \, ,
\label{eq:gflow_latt}
\end{equation}
where
\begin{equation}
\nabla^a f(U) = \lim_{\epsilon \to 0} \frac{f\left(\exp(i\epsilon t^a)\, U\right) - f(U)}{\epsilon}
\, .
\label{eq:nabla}
\end{equation}
The functional derivative is evaluated for a general improved action with the result
\begin{equation}
t^a \nabla^a_{x,\mu} S[U] = -\frac{\beta}{2N}\, Q_\mu(x) \, , 
\label{eq:derivRes}
\end{equation}
where
\begin{equation}
Q_\mu(x) = \frac{i}{2}\left[\left(\Omega_\mu(x) - \Omega_\mu^\dagger(x)\right)
- \frac{1}{N}\,{\rm tr}\left(\Omega_\mu(x) - \Omega_\mu^\dagger(x)\right)\right] \, ,
\label{eq:Qmu}
\end{equation}
is the Hermitian traceless matrix familiar from stout-link smearing \cite{Morningstar:2003gk},
built from the link products
\begin{equation}
\Omega_\mu(x) = U_\mu(x) \sum_{\nu \ne \mu} \Sigma_{\mu\nu}(x),
\label{eq:Omega_def}
\end{equation}
with the staples $\Sigma_{\mu\nu}(x)$ associated with the link $U_\mu(x)$, defined by the
local action 
\begin{equation}
S_{\rm local}[U_\mu(x)] = - \frac{\beta}{N} \, \sum_{\nu \ne \mu} {\rm Re}\,{\rm tr}\left[U_\mu(x)\,
  \Sigma_{\mu\nu}(x)\right] \, , 
\label{eq:stapleDef}
\end{equation}
for SU($N$) gauge field theory.
Thus
\begin{equation}
\frac{d}{dt_g} U_\mu(x;t_g) = i\, Q_\mu(x;t_g)\, U_\mu(x;t_g) \, .
\label{eq:linkFlow}
\end{equation}
Gradient flow is a continuous realization of stout-link smearing. 
Numerical integration of the gradient flow is performed via
\begin{equation}
U_\mu(x;t_g+\rho) = \exp\!\left(i \rho\, Q_\mu(x;t_g)\right)\, U_\mu(x;t_g) \, .
\label{eq:integration}
\end{equation}
The finite transformation generated by gradient flow after flow time $t_g = n_s\, \rho$ is
equivalent to stout-smeared links after $n_s$ sweeps for sufficiently small increments $\rho$.

Our implementation of the gradient flow uses the standard four-dimensional stout-link smearing
algorithm of Ref.~\cite{Morningstar:2003gk}, annealed to effectively update one link at a time,
{\it i.e.}\ sequential single-link updates within each sweep, as opposed to a simultaneous update
of all links.  Our parallel algorithm implements this with link updates uniformly spread throughout
the lattice volume \cite{Bonnet:2000db}. Throughout this quantitative analysis, we use a small
isotropic smearing parameter of $\rho =0.005$. The gradient flow time is $t_g = n_s\, \rho$, where
$n_s$ denotes the number of sweeps of stout-link smearing over the lattice. In conveying the extent
of smearing in the following, we refer to the number of sweeps, $n_s$, or the associated gradient
flow time, $t_g$, interchangeably.

\subsection{Gradient flow actions}
\label{subsec:gfActions}

Here we present how the optimal improved gradient-flow action and its associated optimal flow time
is determined. Historically the simple plaquette gauge-field action was used to define the gradient
flow. However, more recently improved gauge-field actions are being considered
\cite{Tanizaki:2024zsu}. These include the ${\cal O}(a^2)$-improved Symanzik gauge action
\cite{Symanzik:1983gh}, the well known Iwasaki action \cite{Iwasaki:1983iya} and the ``Doubly
Blocked from Wilson action in two-coupling space'' (DBW2) action \cite{Takaishi:1996xj}.

In addition to these we will consider an overimproved \cite{GarciaPerez:1993lic} action which we
will refer to as the Moran \cite{Moran:2008ra} action. This particular action was designed to
stabilize instantons under smoothing. Whereas the Symanzik improved action will shrink
instantons under smoothing, the Iwasaki and DBW2 actions will grow instantons under smoothing. For
the DBW2 action, we anticipate rapid growth in the size of topological objects within the gauge
fields. While we anticipate the Moran action may prove to be optimal for the gradient flow, we also
introduce an intermediate gauge field action that bridges the gap between the Moran and Iwasaki
actions.

All of these gauge field actions may be succinctly described in a single equation. Here we closely
follow the notation of Ref.~\cite{Moran:2008ra}. They decided to use the traditional combination of
plaquettes and rectangles, defined as
\begin{eqnarray}
    S(\epsilon) &=& \beta \sum_x \sum_{\mu > \nu} \bigg[
    \frac{5-2\epsilon}{3} ( 1 - P_{\mu\nu}(x) ) \nonumber \\
&&    - \frac{1-\epsilon}{12} \big( ( 1 - R_{\mu\nu}(x) ) + ( 1 -
    R_{\nu\mu}(x) ) \big) \bigg] \,.
  \label{eqn:overimpaction}
\end{eqnarray}
Here $P_{\mu\nu}$ is the standard plaquette in the $\mu$-$\nu$ plane. It includes taking $1/2$ of
the real part of the trace of the product of the SU(2) links around the elementary plaquette
located at space-time site $x$.  Similarly $R_{\mu\nu}$ refers to link products around $1\times 2$
rectangles.  The consideration of both $R_{\mu\nu}$ and $R_{\nu\mu}$ ensures both rectangle
orientations are considered.

Note that the $\epsilon$ parameter in Eq.~(\ref{eqn:overimpaction}) has been designed such that
$\epsilon = 1$ gives the standard Wilson action, and $\epsilon = 0$ results in the Symanzik ${\cal
  O}(a^2)$-improved action \cite{Moran:2008ra}. More generally, the action may be written as
\begin{eqnarray}
    S(\epsilon) &=& \beta \sum_x \sum_{\mu > \nu} \bigg[
    C_p \, ( 1 - P_{\mu\nu}(x) ) \nonumber \\
&&    - C_r \, \big( ( 1 - R_{\mu\nu}(x) ) + ( 1 -
    R_{\nu\mu}(x) ) \big) \bigg] \,.
  \label{eqn:oimpgen}
\end{eqnarray}
For  given $C_r$
and $C_p$ coefficients, 
the $\epsilon$ parameter is given by
\begin{equation}
\label{eq:eps}
 \epsilon = \frac{1-20\, \frac{C_r}{C_p}}{1-8\, \frac{C_r}{C_p}} \quad.
\end{equation}
Table~\ref{tab:gradFlow} summarizes the $\epsilon$ values associated with the different gradient
flow actions under consideration. 

Here we emphasize how the Symanzik improved action with $\epsilon = 0$ has an ${\cal O}(a^4)$ error
that causes instantons to shrink in size under smoothing algorithms.  Moran {\it et al.}
\cite{Moran:2008ra} introduced the epsilon parameter to over improve the action and change the sign
of the error term for instantons. By using a small negative value for $\epsilon$ the size of
instantons is stabilized under smoothing. 

These observations lead to some concern in using the Iwasaki or DBW2 actions within the gradient
flow algorithm. Both actions have negative $\epsilon$ values, but the magnitude of the values is
relatively large.  The concern is that large discretization errors in the lattice action will
enable the gauge field action to be reduced by growing the size of instantons. Thus we are
concerned that the Iwasaki and DBW2 actions may not be ideal for preserving the topological
structure of the gauge fields.  Indeed the renormalization issues that lead to these actions will
be removed under gradient flow, making their selection for smoothing poorly motivated.

To fill the gap between the Moran and Iwasaki values for $\epsilon$ in Table~\ref{tab:gradFlow}, we
also include an intermediate action with $\epsilon=-1$.

\begin{table}[tbp]
\caption{Plaquette ($ C_p $) and rectangle ($ C_r $) coefficients as defined in
  Eq.~(\ref{eqn:oimpgen}) for different gradient flow actions and their corresponding $\epsilon$
  values.
}
\label{tab:gradFlow}
\centering
\begin{ruledtabular}
\begin{tabular}{c c c c} 
Gradient flow & $ C_p $ & $ C_r $ & $ \epsilon $ \\
\hline
\noalign{\smallskip}
Wilson & 1.0000 & 0.0000 & 1 \\
Symanzik & 1.6667 & 0.0833 & 0 \\
Moran & 1.8333 & 0.1042 & $-$0.25 \\
Intermediate & 2.3333 & 0.1667 & $-$1 \\
Iwasaki & 3.6480 & 0.3310 & $-$2.972 \\
DBW2 & 12.2704 & 1.4088 & $-$15.9056 \\
\end{tabular}
\end{ruledtabular}
\end{table}

\subsection{Smoothing criteria}
\label{smoC}

The criterion used in this research to determine the optimal smoothing is the dimensionless quantity defined as
\begin{equation}
\label{eq:F}
 {\cal F} = t^2_g \, \langle | q(x) | \rangle_x \quad.
\end{equation}
Here $t_g$ is the gradient flow time and $\langle | q(x) | \rangle_x$ is the magnitude of the
topological charge density averaged over the lattice sites $x$. 
Recalling that the topological charge density has units of GeV$^4$, and the gradient flow has a
physical smoothing distance $d=\sqrt{8t_g}$ \cite{Luscher:2010iy}, this gradient flow measure,
${\cal F}$, is dimensionless.

It can be deduced from Eq.~(\ref{eq:F}) that the derivative of ${\cal F}$ is linear in $t_g$ when
$\langle | q(x) | \rangle_x$ is constant.  Reductions in $\langle | q(x) | \rangle_x$ will flatten
the curve, so early in the smearing, flattening of the first derivative (i.e.negative values for
the second derivative) is a sign that the gradient flow is efficiently removing topological charge
density in a local manner. This is the optimal phase where the gauge fields are smoothed with
minimal changes in the long-distance field structure. Thus, there is no need to stop smoothing
until the first derivative approaches its long term general trend. In the long term general trend,
$\langle | q(x) | \rangle_x$ approaches a constant and we enter a phase where opposite-charge
topological objects are migrating toward each other and eventually annihilating. The process of
annihilation will induce ripples into the dimensionless quantity ${\cal F}$ as rapid changes in the
topological charge density magnitude take place. In this phase, smoothing is changing the structure
of the fields.

\begin{figure*}[tb]
\centering
\includegraphics[width=0.495\textwidth]{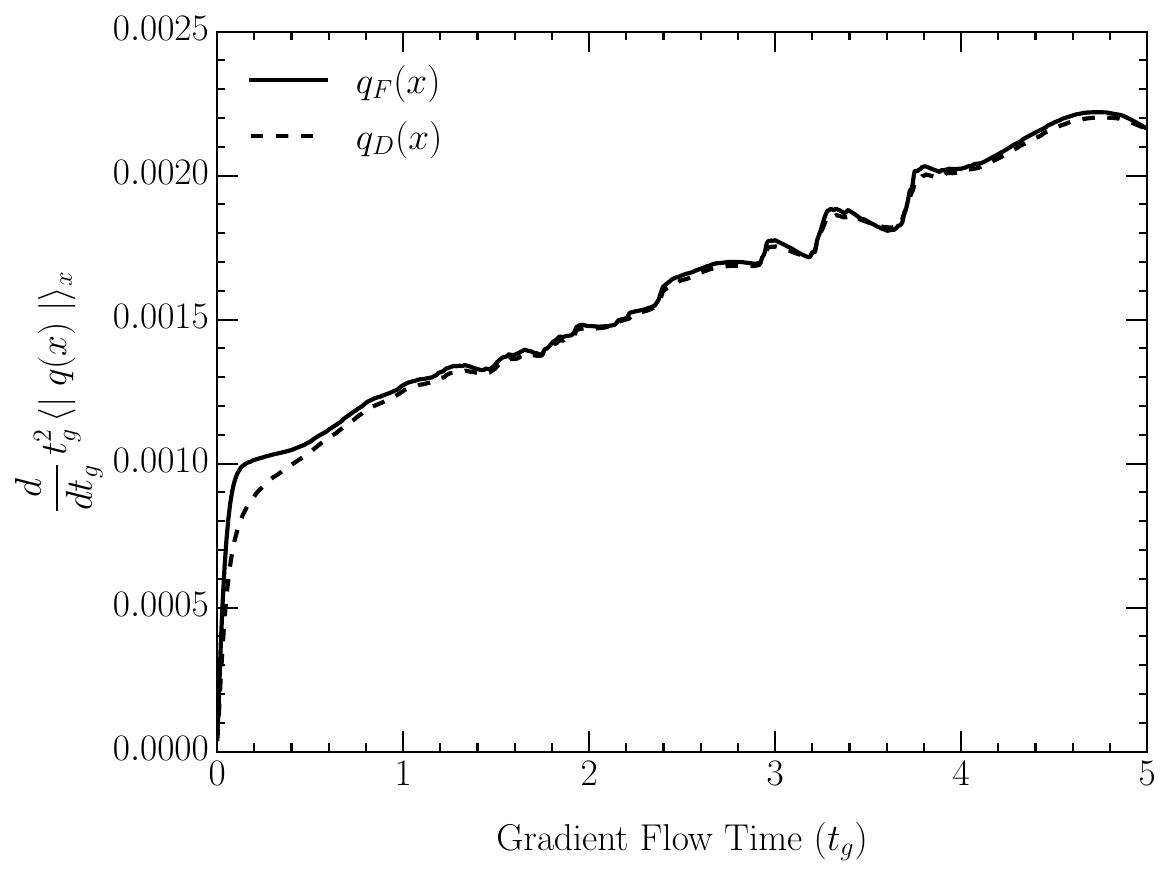}
\includegraphics[width=0.495\textwidth]{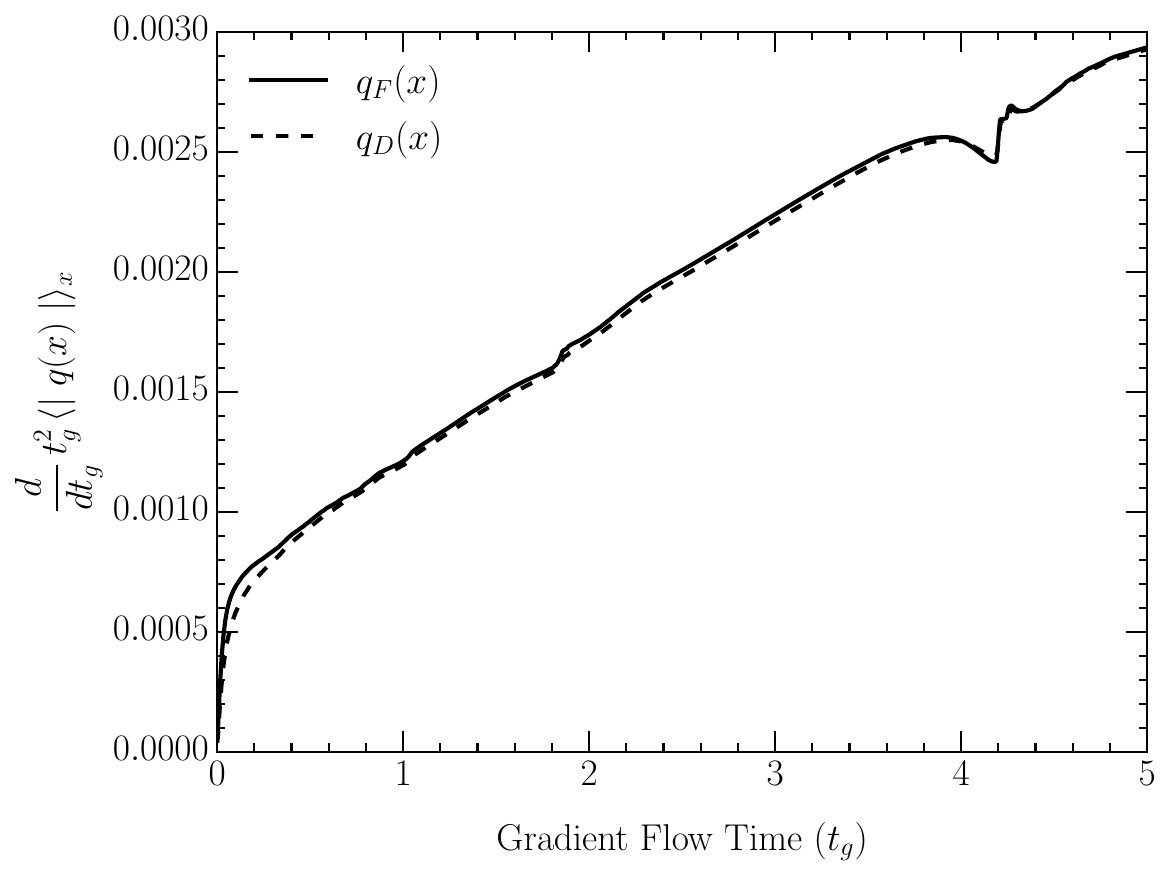}
\includegraphics[width=0.495\textwidth]{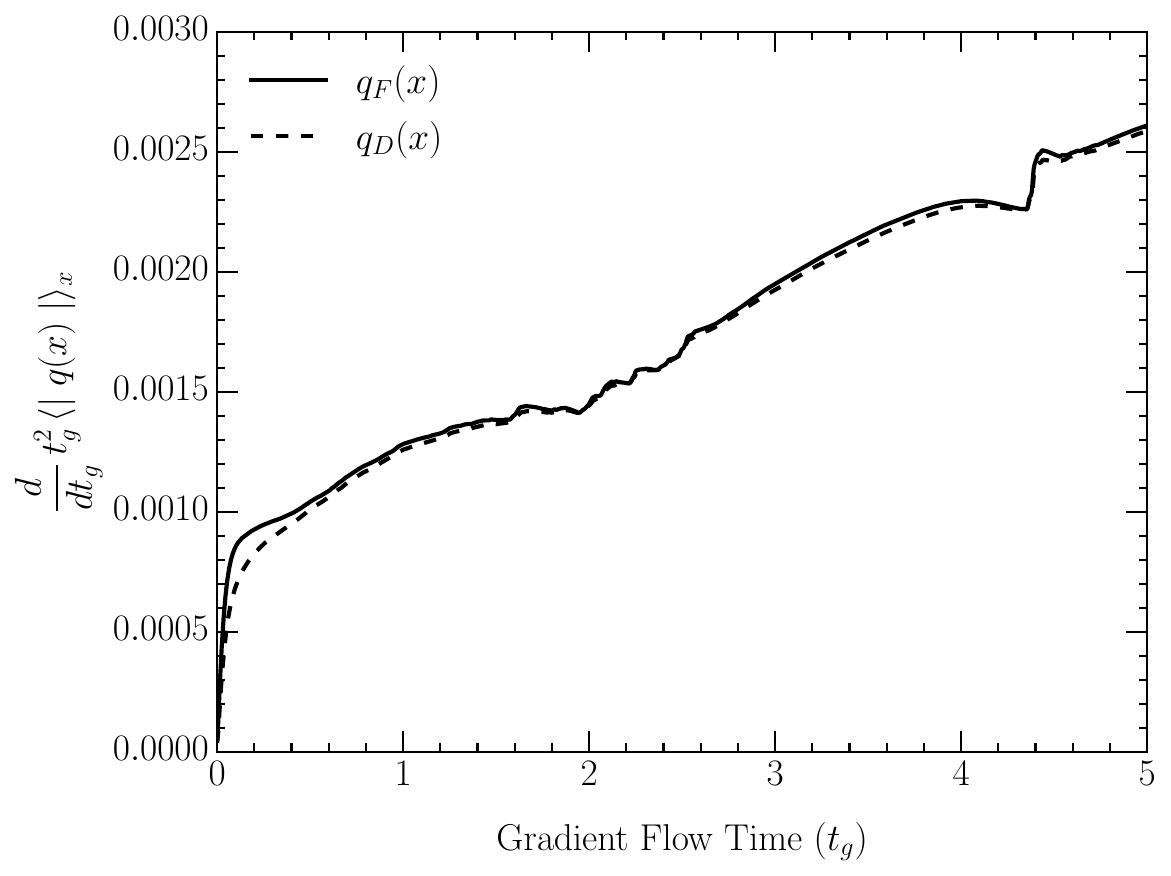}
\includegraphics[width=0.495\textwidth]{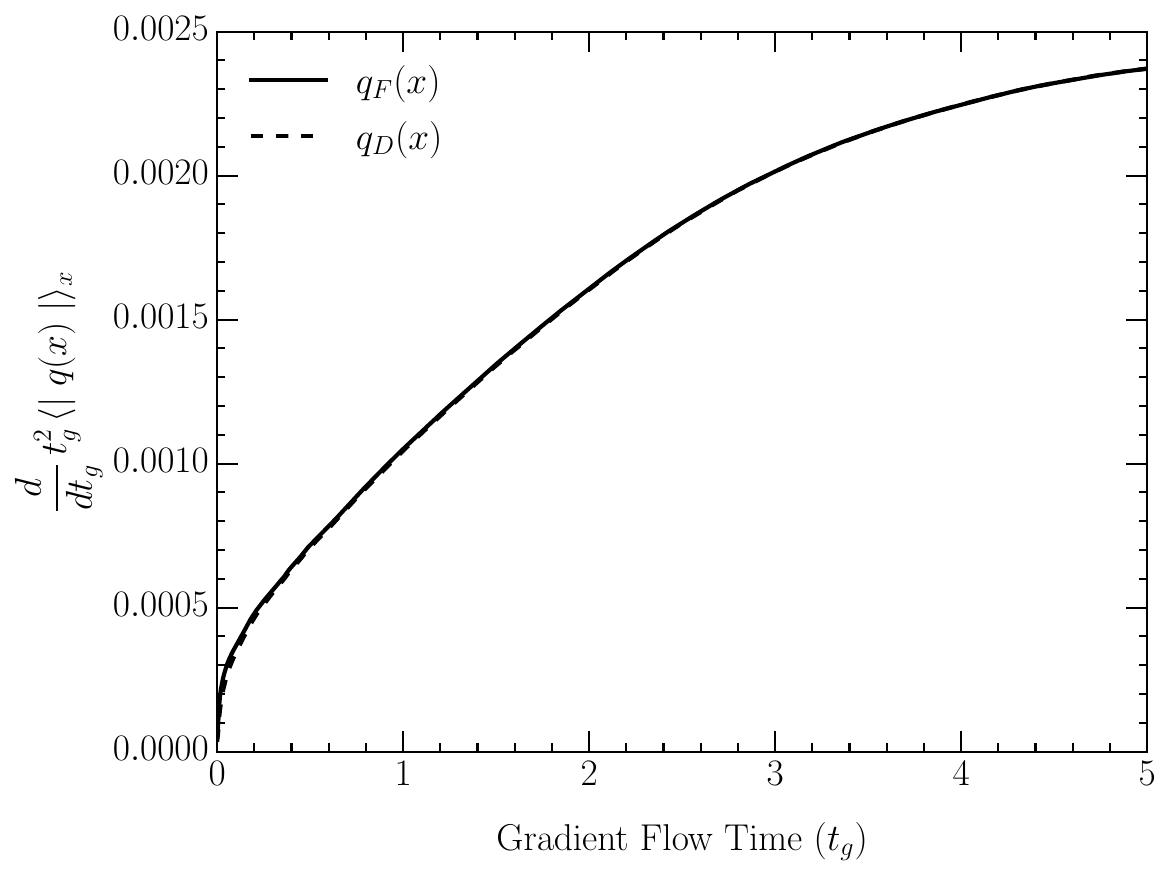}
\caption{Representative plots of one configuration with zero chemical potential illustrating
  the first derivative of ${\cal F}$ for the four gradient flow actions: Moran (top left),
  Intermediate (bottom left), Iwasaki (top right) and DBW2 (bottom right).
}
\label{fig:Lat32_smoC}
\end{figure*}

We commence our analysis by considering a survey of chemical potential, $\mu$, and source term
strengths, $j$, on the fine ensemble ($\beta=2.1, \kappa=0.1577$).  These lattices are
relatively smooth in comparison to the coarse ensemble lattices that
will be used later for examining the
chemical potential dependence of the ground-state field structure.

In calculating the topological charge density for ${\cal F}$, we consider both of the improved
measures discussed in Sec.~\ref{subsec:impOps}. In the following discussion our attention will
be on the topological charge density obtained through the improvement of the lattice field strength
tensor. Results from this measure will be compared with the direct improvement of the topological
charge density. As we will see, results from the improved lattice field strength tensor will show
more structure in the gradient flow, and this is helpful in deriving a stopping criterion for the
threshold amount of gradient flow.

Our survey of $\mu$ and $j$ values revealed little sensitivity, and the selection of any parameter
values may be regarded as representative.  Figure~\ref{fig:Lat32_smoC} shows how the gradient flow
measure ${\cal F}$ changes for the four gradient flow actions under consideration. In this case we
have selected $\mu = 0$.  All four plots show a sharp rise in the first derivative of ${\cal F}$ at
the earliest flow times. Setting the DBW2 action aside, the other three actions then show a
flattening of the first derivative. As described earlier, this indicates significant reduction in
the topological charge density magnitude. It is here that one aims to continue smoothing as the
gradient flow is efficiently removing local fluctuations. However, this flattening is seen to end
fairly rapidly. The Moran gradient flow leaves the flattening phase at $t_g\approx 0.6$, the
Intermediate gradient flow departs at $t_g\approx 0.5$, and the Iwasaki gradient flow leaves at
$t_g\approx 0.4$.

At this point, one enters the long term trend seen over a wide range of gradient flow time. It is
here that one becomes concerned that the structure of the gauge fields is being significantly
modified through the gradient flow and thus this region is to be avoided. The DBW2 gradient flow
never quite flattens out the derivative curve, possibly because the instantons grow very quickly
and start annihilating immediately after the flow begins.

There is another important observation to be drawn from Fig.~\ref{fig:Lat32_smoC}. As the magnitude
of $\epsilon$ increases, the gradient flow criterion, ${\cal F}$, becomes increasingly smooth. This
is in accord with expectations as emphasized earlier. Large values of $\epsilon$ will cause the
instantonlike topological objects to grow in size. For large gradient flow time, the objects are
likely to become quite large and overlapping. This contrasts the Moran action which is designed to
leave the topological structures intact and unaltered under the smoothing algorithm. In this case
one can imagine well-defined objects moving across the lattice and occasionally finding neighbors
to annihilate with. It is here that large changes in the topological charge density will take place
and we will observe ripples in the smoothing criterion, ${\cal F}$.

In light of this, it is tempting to discard all actions other than the Moran action. However, the
Iwasaki action is well known; it is currently being used within the literature to perform the
gradient flow; it does show the fluctuations associated with well-defined objects. Therefore, we
feel it will be informative to investigate this gradient flow action further. However, the DBW2
action is not a useful gradient flow action for this research.

Therefore, Moran and Iwasaki are selected to study QC${}_2$D further. The Intermediate gradient
flow is dropped since it shows no advantage over the Moran action.

\subsection{Gradient flow effect on coarse lattices}
\label{coLat}

\begin{figure*}[tb]
\centering
\includegraphics[width=0.495\textwidth]{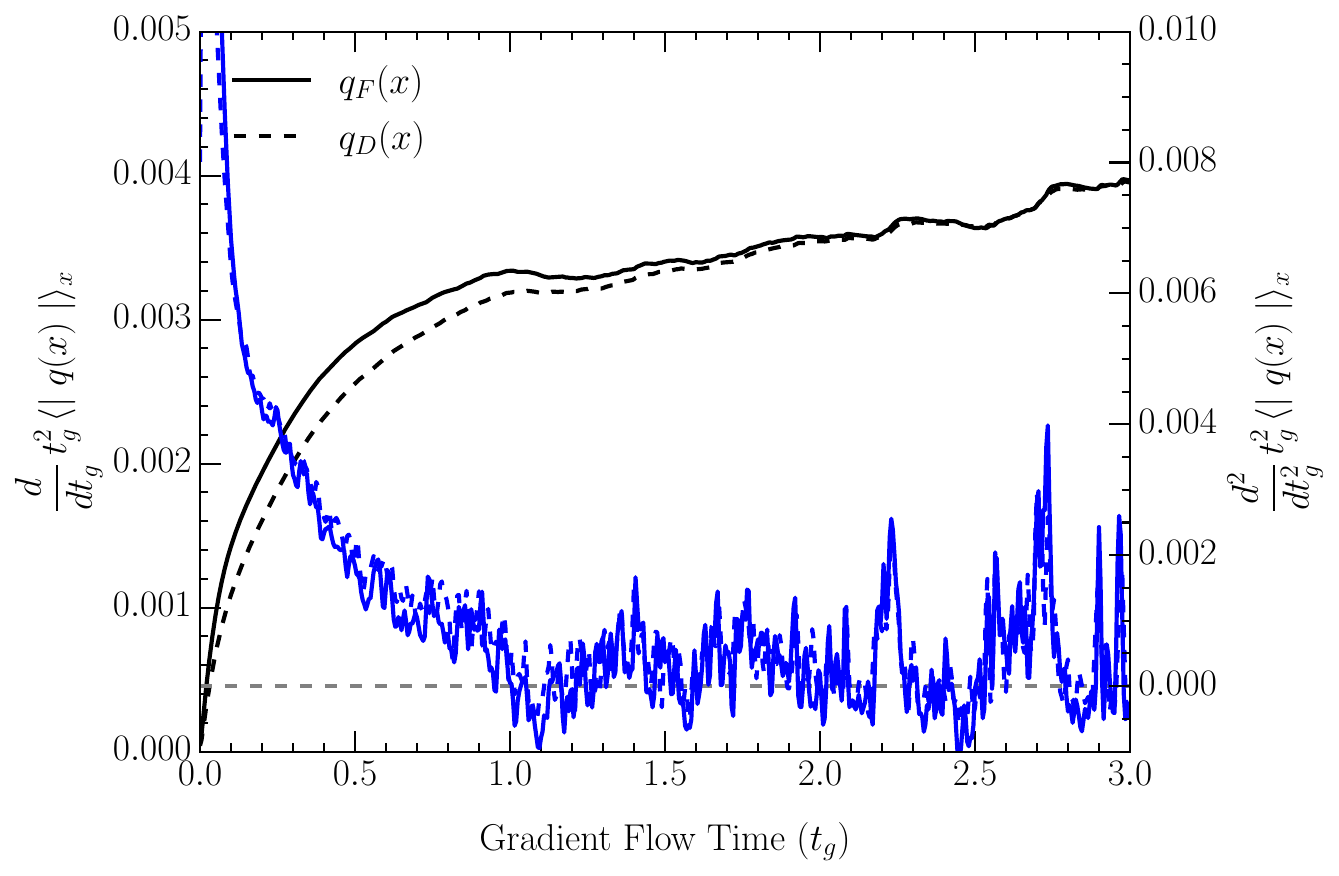}
\includegraphics[width=0.495\textwidth]{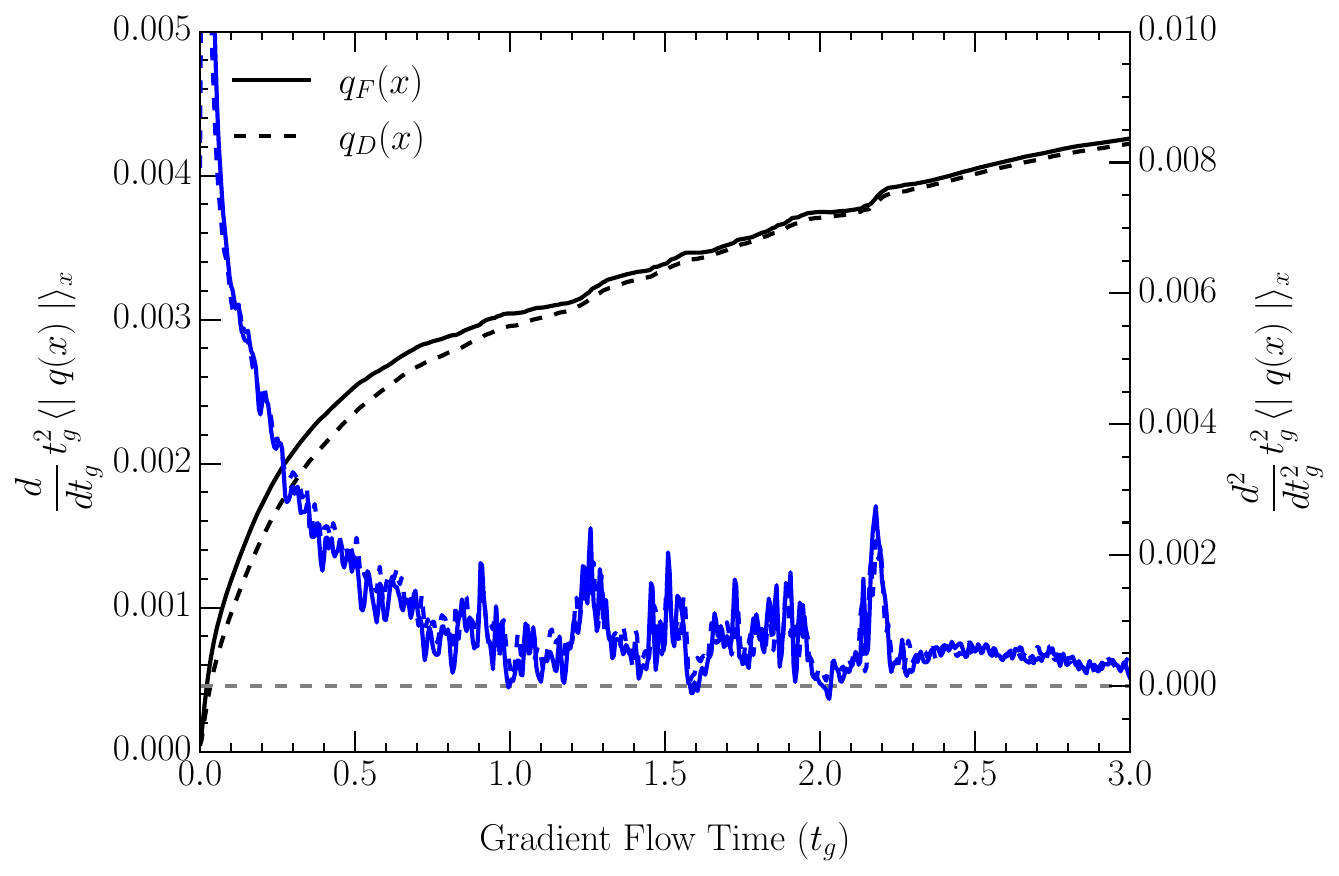}
\caption{Representative gradient flows for a coarser $16^3 \times 24$ lattice with $a\mu = 0.25$
  and $a j = 0.2$.  Both the first derivative (black lines) and second derivative (blue lines)
  of ${\cal F}$ are illustrated for Moran (left) and Iwasaki (right) gradient flows.
}
\label{fig:Lat24_smoC}
\end{figure*}

The SU(2) configurations available for the study of the finite-density chemical-potential impact on
the ground-state field structure are the coarse ensemble with $\beta=1.9$ and
$\kappa=0.1680$, providing a lattice spacing of $a=0.178(6)$ fm
\cite{Cotter:2012mb}. Representative plots for the gradient flow results analogous to
Fig.~\ref{fig:Lat32_smoC} are provided in Fig.~\ref{fig:Lat24_smoC} for our two preferred
gradient-flow actions. The more coarse nature of the lattice spacing presents new challenges in
identifying the threshold regime for the gradient flow time. Thus, the second derivative of ${\cal
  F}$ is added in the plots to aid in making a precise determination of the flattening of the first
derivative curve on these coarser lattices.

As seen in Fig.~\ref{fig:Lat24_smoC}, the Moran gradient flow enters the long-term trend at
$t_g\approx 1.0$ and the Iwasaki gradient flow enters a little earlier at $t_g\approx 0.8$. These
flow times correspond to 200 (Moran) and 160 (Iwasaki) sweeps of stout-link smearing with
$\rho=0.005$, respectively.

\subsection{Highly improved topological charge operators}
\label{ops}

To further examine the nature of the gradient flow, we examine the difference between the two
highly improved estimates of the topological charge density presented in
Sec.~\ref{subsec:impOps}. Recall that one measure first improves the lattice field strength
tensor and then constructs the topological density from this improved operator; we refer to this
estimate as $q_F(x)$. The other measure directly improves the topological density and is referred
to as $q_D(x)$. While both schemes provide an ${\cal O}(a^4)$-improved measure of the topological
charge density, the ${\cal O}(a^6)$ errors are different. Our goal is to understand where
these differences are sufficiently small such that we can measure the chromoelectric and
chromomagnetic field strengths in an accurate manner.

\begin{table}[tb]
\caption{Performance of the two preferred gradient flow actions on the fine $16^3 \times 32$
  lattices with $ \beta = 2.1$ and $ \kappa = 0.1577$ and the coarse $16^3\times24$ lattices with
  $\beta=1.9$ and $\kappa=0.1680$. The gradient flow is quantified by examining the differences
  between the two ${\cal O}(a^4)$-improved topological charge density measures. The relative error
  defined in Eq.~(\ref{eq:relErr}) at the 120th sweep (Error$_{120}$) is indicated in percent, and
  the number of sweeps required to reduce the error to less than 10$\%$ (Sweeps for $<$10$\%$) are
  indicated. The values quoted are the median values observed in a small set of test runs as
  described in the text, with the uncertainties (in the last digit) given by the deviation of the
  upper and lower values from the median.}
\label{tab:errorFlow}
\centering
\begin{ruledtabular}
\begin{tabular}{c c c c c c} 
Flow & $\beta$ & $\kappa$ & $a \mu$ & Err$_{120}$ & Sweeps for $<$10$\%$ \\
\hline
\noalign{\smallskip}
\multirow{3}{4em}{Moran} &\multirow{3}{2em}{2.1}&
\multirow{3}{3em}{0.1577}&
0.0 & 10.5\err{2}{3} & 127\err{4}{4}  \\ 
& & & 0.4 & 10.3\err{1}{4} & 124\err{2}{5}  \\
& & & 0.7 & 10.6\err{3}{2} & 129\err{4}{2}  \\
\hline
\noalign{\smallskip}
\multirow{3}{4em}{Iwasaki} &\multirow{3}{2em}{2.1}&
\multirow{3}{3em}{0.1577}&
0.0 & 7.4\err{2}{3} & 85\err{2}{3}  \\ 
& & & 0.4 & 7.1\err{4}{3} & 83\err{1}{3}  \\
& & & 0.7 & 7.5\err{2}{1} & 86\err{3}{2}  \\
\hline\hline
\multirow{3}{4em}{Moran} &\multirow{3}{2em}{1.9}&
\multirow{3}{3em}{0.1680}&
0.0 & 15.0\err{1}{3} & 218\err{5}{5}  \\ 
& & & 0.25 & 14.7\err{1}{1} & 211\err{1}{0} \\
\hline
\noalign{\smallskip}
\multirow{3}{4em}{Iwasaki} &\multirow{3}{2em}{1.9}&
\multirow{3}{3em}{0.1680}&
0.0 & 11.4\err{1}{1} & 145\err{2}{2}    \\ 
& & & 0.25 & 11.2\err{0}{1} & 141\err{0}{1}  \\
\end{tabular}
\end{ruledtabular}
\end{table}

We introduce a relative error defined by
\begin{equation}
\label{eq:relErr}
  R_E= \frac{\langle | q_F(x) | \rangle_x - \langle | q_D(x) | \rangle_x}{\frac{1}{2} \left
    (\langle | q_F(x) | \rangle_x + \langle | q_D(x) | \rangle_x \right )}\, ,
\end{equation}
where we once again work with the average magnitudes of topological charge density obtained from
the improved lattice field strength tensor operator and the direct topological density
operator.  Table~\ref{tab:errorFlow} tabulates the relative errors between
these operators at $120$ sweeps of Moran and Iwasaki gradient flows for both the $\beta=2.1$ and
$\beta=1.9$ ensembles respectively. It also records the number of sweeps taken for the relative
error to drop below $10 \%$. Drawing on this last criterion, in concert with the results of
Sec.~\ref{coLat} focusing on the dimensionless flow parameter, ${\cal F}$, the optimal sweep
numbers for the $\beta=1.9$ ensemble are determined as 200 (Moran)
and 150 (Iwasaki). These sweep numbers correspond to gradient flow times of 1.00 (Moran) and 0.75
(Iwasaki).

We have also considered the Euclidean action remaining for each of these flows using an action
constructed from the improved field strength tensor. We find 200 sweeps of Moran gradient flow
corresponds to 154 sweeps of the Iwasaki flow, confirming our expectations. Thus, our
flow selections are providing a similar field structure.

\subsection{Renormalized operators}
\label{RO}

To obtain the continuum values for the magnetic and electric gluon condensates, $\braket{E^2}$ and
$\braket{B^2}$, our results obtained at finite gradient flow time need to be renormalized
\cite{Luscher:2011bx}.  In general terms, the process consists of describing the flow dependence of
an operator, ${\cal O}(x)$, via an operator product expansion for small flow times, $t_g$
\begin{equation}
{\cal O}(x,t_g) = \sum_i^n C_i(t_g)\, {\cal O}_i(x) + {\cal O}_h(t_g) \, ,
\end{equation}
where the $C_i(t_g)$ are perturbatively calculable $t_g$-dependent matching (Wilson) coefficients
connecting flowed operators ${\cal O}(x,t_g)$ to the desired renormalized local operators ${\cal
  O}_i(x)$ in a continuum scheme. The term ${\cal O}_h(t_g)$ captures higher-order terms of the OPE
not included in the sum over $i$.

There is a window in flow time that must be met in order to carry out this renormalization
procedure.  The physical flow diffusion radius \cite{Luscher:2010iy} $d = \sqrt{8 t_g\, a^2}$ must
satisfy two competing requirements.

First, $d$ must remain small, such that the asymptotic expansion of perturbation theory is reliable.
The flow damps momentum modes above the scale $\mu = 1/d$, where QCD is weakly coupled; the effect
of these suppressed short-distance modes is absorbed into the perturbatively calculable Wilson
coefficients evaluated at the scale $\mu$. Asymptotic freedom ensures that the strong coupling
$\alpha_s(\mu)$ is small for large $\mu$.
Thus, the scale $\mu$ must be large enough that the perturbative expansion of the Wilson
coefficients, $C_i(t_g)$, can be truncated at fixed loop order, a condition governed by
$\alpha_s(\mu)$.

A second requirement demanding small flow times has its origin in the neglected higher-order
operators in the operator product expansion, ${\cal O}_h(t_g)$.  These contributions must be
suppressed sufficiently to enable a simple $t_g$-dependent ansatz to capture their contribution.
This second condition follows from dimensional counting, since operators of dimension $d_j >
d_{{\cal O}_n}$ enter with coefficients suppressed by $t_g$ raised to the power ${(d_j - d_{{\cal
      O}_n})/2}$, independent of the coupling.

The competing requirement in flow time is that the physical diffusion scale must be large compared
to the lattice spacing so that discretization artifacts in the flowed operator, ${\cal O}(x,t_g)$,
are suppressed.  To this point in our investigation, we have determined the threshold gradient flow
time where vacuum expectation values of ${\cal O}(x,t_g)$, namely $\braket{q(x,t_g)}$,
$\braket{E^2(x,t_g)}$, and $\braket{B^2(x,t)}$, are under control.  Herein, the somewhat coarse
lattice spacing has presented some challenges.  However, our use of ${\cal O}(a^4)$-improved
operators has enabled us to obtain results at a modest flow time of $t_g \sim 1$.  At this flow
time, the physical dimensionful diffusion scale of the gradient flow provides $d = \sqrt{8}\, a$, a
scale reasonably larger than the lattice spacing, providing another perspective on the required
threshold flow time.

All of the aforementioned flow-time requirements encourage the consideration of small lattice
spacings, particularly to meet the demands of the operator product expansion.  However, with our
physical lattice spacing of $a = 0.178(6)$ fm, the diffusion radius is $d = 0.50(2)$ fm, deep in
the nonperturbative regime and corresponding to a scale of merely $\mu = 0.4$ GeV.
Since the minimum diffusion radius required to determine ${\cal O}(x,t_g)$ lies beyond the scale
within which perturbative matching is trustworthy, a flow-time window that simultaneously satisfies
both requirements does not exist on this ensemble, and we therefore cannot proceed further.

This tension between the flow requirements is not unique to our lattices. The renormalization
process has been carried out in Ref.~\cite{Datta:2015bzm} using the Wilson flow to study the SU(3)
deconfinement transition.  There the renormalization of the magnetic and electric gluon condensates
was carried out at leading order.  Using the physical critical temperature of the deconfinement
transition in SU(3) Yang-Mills theory of $T_c = 290$ MeV \cite{Borsanyi:2022xml}, one can infer
that their lattice spacings are significantly smaller than that considered here.  For example, on
their $N_t = 8$ lattices, the spacings are in the range 0.04 to 0.09 fm.

Even in this optimistic case, the authors concluded that extracting the renormalized gluon
condensates was difficult.  They found a flow-time dependence that was not accounted for in the
leading-order perturbative renormalization constant.  With the relevant two-loop matching
coefficients now derived and under study \cite{Harlander:2018zpi,Iritani:2018idk,Taniguchi:2020mgg}, it
would be timely to invest the significant resources required to repeat this analysis with
renormalized operators in a next-generation calculation.

\section{Finite-density modifications of the ground-state fields}
\label{sec:survey}

\subsection{Quantitative results}
\label{QR}

The conventional Wilson gauge action with two Wilson-fermion flavors is used in
this research to investigate QC$_2$D. As presented in Sec.~\ref{sec:gaugeFields}, the fermion
action is constructed by a gauge- and isosinglet diquark source term which lifts the low-lying
eigenvalues of the Dirac operator while allowing for a controlled study of diquark condensation
\cite{Cotter:2012mb}. Recall the fermion action is 
\begin{equation}
S_Q+S_J=\sum_{i=1,2}\bar\psi_iM\psi_i
 + \kappa j[\psi_2^{\rm tr}(C\gamma_5)\tau_2\psi_1-{\rm H.c.}],
\label{eq:Slatt1}
\end{equation}
with
\begin{align}
M_{xy}=\delta_{xy}-\kappa\sum_\nu&\Bigl[(1-\gamma_\nu)e^{\mu\delta_{\nu0}}U_\nu(x)\delta_{y,x+\hat\nu}\nonumber\\
&+(1+\gamma_\nu)e^{-\mu\delta_{\nu0}}U^\dagger_\nu(y)\delta_{y,x-\hat\nu}\Bigr].
\label{eq:Mwils1}
\end{align}

We begin by ignoring a possible source term dependence in the results,
and compare results obtained with $aj=0 (\mu=0), aj=0.02 (a\mu=0.25)$
and $aj=0.04 (a\mu>0.25)$.  100--200 gauge configurations from the
coarse ensemble are used at each chemical potential for this analysis.
The dotted line in all the plots represents the $m_\pi/2$ phase boundary.

\begin{figure}[tb]
\centering
\includegraphics[width=\columnwidth]{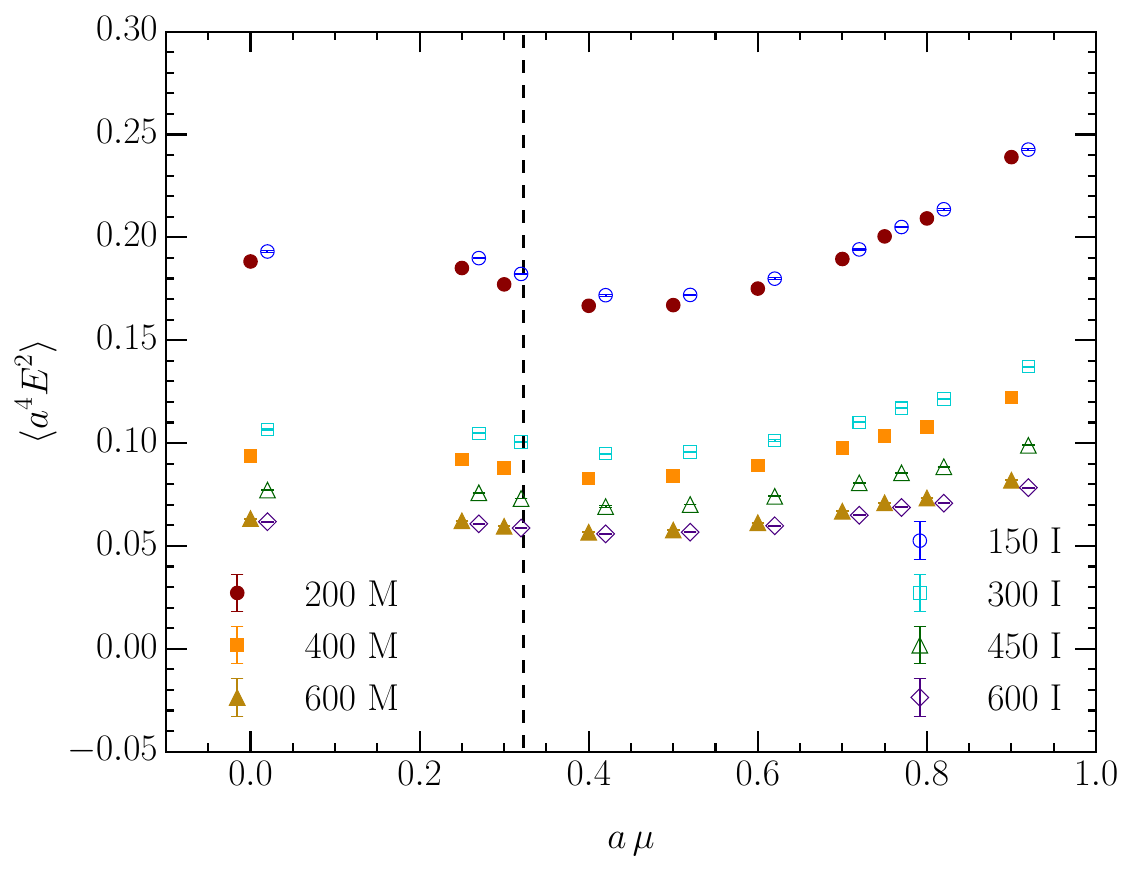}
\caption{Dependence of the squared chromoelectric field strength on the chemical potential. Results
  are illustrated for the Moran (M) and Iwasaki (I) gradient flows at several levels of smearing as
  indicated by the number of smearing sweeps applied with $\rho = 0.005$ in the legend. The points
  depicting the Iwasaki gradient flow in the plot are shifted by $a\mu=0.02$ for clarity.
}
\label{fig: Esq}
\end{figure}

\begin{figure}[tb]
\centering
\includegraphics[width=\columnwidth]{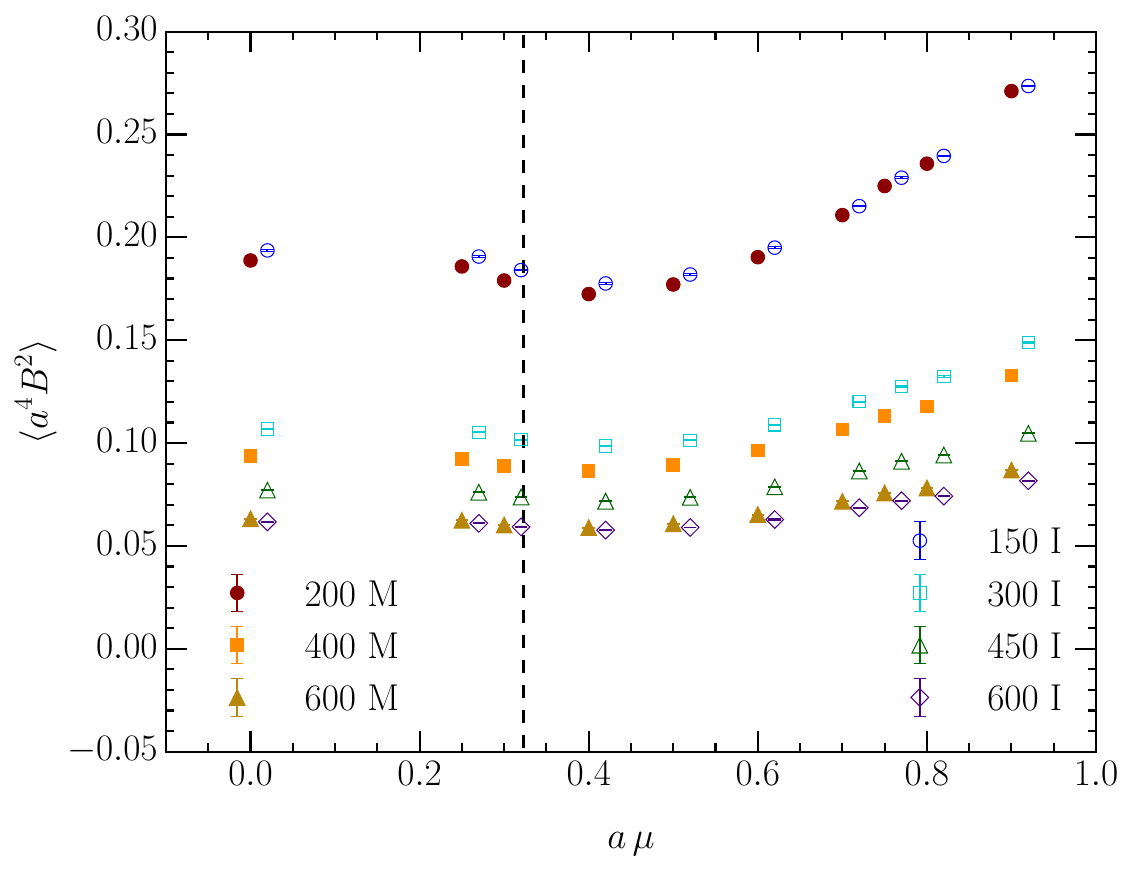}
\caption{Dependence of the squared chromomagnetic field strength on the chemical potential. Results
  are as illustrated in Fig.~\ref{fig: Esq}.
}
\label{fig: Bsq}
\end{figure}

Figure~\ref{fig: Esq} shows results for the squared chromoelectric field strength dependence on the
chemical potential. Noting that increased gradient flow suppresses the chromo-electromagnetic
fields, the best signals are obtained with fewer smoothing sweeps, where a clear suppression of the
field strengths is manifest above $a\mu =0.4$. The gauge field structures are suppressed at higher
levels of smearing. However, all the results show the same trend; the chromoelectric field
strength drops near $m_\pi/2$, but it regains strength at higher chemical potentials and becomes
much stronger at the highest chemical potentials considered.

The same trend is observed for the squared chromomagnetic field strength, as presented in Fig.~\ref{fig:
  Bsq}. While this research illustrates results for very large values of the lattice chemical
potential, $a\mu$, lattice artifacts are likely problematic for the largest values considered. With
this in mind, we regard the most interesting region to be $a\mu \leq 0.7$. This is in
  agreement with earlier results, which have found that the Polyakov loop increases from zero at
  $a\mu\gtrsim0.7$, independent of the lattice spacing \cite{Boz:2019enj}.

It is interesting to note how both the chromoelectric and chromomagnetic field strengths drop near
$a\mu \simeq 0.3$ in advance of the anticipated phase transition at $m_\pi/2$. While it is noted
that the consideration of a finite volume system necessarily creates a crossover, the concern is
that the drop in advance of $m_\pi/2$ may be associated with a significant dependence on the
strength of the source term. For example, if the source term acts to suppress the field strength,
then an extrapolation back to zero source term would increase the field strength value, perhaps
maintaining the values observed at zero chemical potential. This possibility is investigated in the
next Section.

\subsection{Source term dependence}
\label{FD}

To investigate the source term dependence of the chromoelectric and chromomagnetic field
strengths, we have used 50 configurations at both $aj=0.02$ and $aj=0.03$ for $a\mu=0.3,\, 0.5,\,
0.7,\, 0.9$ in addition to the 200 configurations at $aj=0.04$ used in the previous Section.  The
chromoelectric and chromomagnetic field strength dependence on the source term is modeled as
\begin{equation} 
\label{eq: slope} 
\langle a^4E^2 \rangle = m_E \, aj + c_E, \quad \langle a^4B^2 \rangle = m_B \, aj + c_B, 
\end{equation} 
where $m$ represents the slope and $c$ represents the $y$ intercept.

\begin{table}[tb]
\caption{Slope parameters $ m_E $ and $ m_B $ describing the source term dependence of the
  squared chromoelectric and chromomagnetic field strengths as defined in Eq.~(\ref{eq: slope}). The
  results shown are obtained after 200 sweeps of Moran and 150 sweeps of Iwasaki gradient flow
  actions.
\vspace{6pt}
}
\label{tab:slope}
\centering
 \begin{ruledtabular}
 \begin{tabular}{c c c c} 
 Gradient flow & $a \mu$ & $ m_E $ & $ m_B $ \\
 \hline
 \noalign{\smallskip}
 \multirow{4}{4em}{Moran (200)} & 0.3 & 0.13 $\pm$ 0.13 & 0.30 $\pm$ 0.13 \\ 
 & 0.5 & 0.29 $\pm$ 0.13 & 0.35 $\pm$ 0.12 \\
 & 0.7 & 0.09 $\pm$ 0.14 & 0.04 $\pm$ 0.13 \\
 & 0.9 & 0.12 $\pm$ 0.12 & 0.07 $\pm$ 0.13 \\
 \hline
 \noalign{\smallskip}
 \multirow{4}{4em}{Iwasaki (150)} & 0.3 & 0.16 $\pm$ 0.14 & 0.30 $\pm$ 0.13 \\ 
 & 0.5 & 0.31 $\pm$ 0.13 & 0.37 $\pm$ 0.12 \\
 & 0.7 & 0.11 $\pm$ 0.14 & 0.04 $\pm$ 0.14 \\
 & 0.9 & 0.09 $\pm$ 0.12 & 0.02 $\pm$ 0.13 \\
 \end{tabular}
 \end{ruledtabular}
\end{table}

As can be seen in Table~\ref{tab:slope}, the slopes are often
consistent with zero, suggesting at most a weak dependence on the
diquark source.  Where a significant
slope is found (notably at $a\mu=0.5$), it is {\em positive}, $m_{E,B}\sim 0.3$.

Considering the source term value, $aj = 0.04$, used for most of the production runs, we see the
correction in extrapolating back to $aj=0$ is to {\em reduce} the value of the squared field
strengths by $\sim 0.01$, a very small correction on the scale of the results presented in
Figs.~\ref{fig: Esq} and \ref{fig: Bsq}. In any event, the source term dependence is not
responsible for the small reduction in the field strengths ahead of the anticipated phase boundary
and we conclude the effect is associated with the finite volume of the lattice.

The small magnitude of these source term corrections contrasts with the very significant
corrections to the diquark condensate reported in Ref.~\cite{Cotter:2012mb}, but is in accord with
the very mild corrections to gluonic quantities found in Ref.~\cite{Boz:2013rca}. Here, the
corrections are an order of magnitude or more smaller than the values themselves. This source term
dependence is also small relative to the smoothing dependence of the results. The magnitude is
similar to the subtle differences between our calibration of the Moran and Iwasaki gradient flow
algorithms. In this context, the correction is not significant and given the large uncertainties in
the source term corrections, we do not consider them further.

\subsection{Estimating the critical chemical potential}
\label{Sig func}

In a finite volume system, sharp first and second order phase transitions are smoothed out to
become crossovers. However, the presence and nature of a phase transition can be inferred through
a systematic study of the behavior of the crossover on the spatial volume of the lattice.

A particularly useful analysis technique is to consider the sigmoid function. The function is fit
to finite volume results and it has been observed that the position of the inflection point is
robust against changes in the volume. In this way, the critical value for the phase transition can
be estimated even in a finite volume analysis.

Therefore, fitting a sigmoid function in the datasets of the squared chromoelectric and
chromomagnetic field strengths obtained will provide an estimate of the chemical potential at
which a transition occurs in infinite volume.

The sigmoid function introduces four parameters which are tuned to describe the lattice results.
\begin{equation}
\label{eq: sig}
  f(\mu) = c_0 + c_1 \arctan(c_2\,(\mu-\mu_c))
\end{equation} 

Here, $\mu_c$ describes the position of the inflection point related to the critical value of the
phase transition. Since there are four parameters in Eq.~(\ref{eq: sig}), at least four data points
are required to find the values. In this Section, the four points selected are $a\mu = 0.00,\ 0.25,\
0.30$ and $0.40$, since these points are the closest to the estimated critical chemical potential
of the phase transition; $a\mu = 0.50$ is not a feasible point to consider since it shows increases
in field strengths from $a\mu = 0.40$, thus distorting the shape of the fitted sigmoid function.

The value $\mu_c$ estimates the chemical potential at which a phase transition occurs in infinite
volume. As seen in Table~\ref{tab:sig}, the uncertainties in the sigmoid fit for the critical
$a\mu$ value increases with the number of sweeps. 
This happens because the higher number of sweeps increasingly distorts and suppresses the field
structures, making the field strengths very similar across the transition regime.  This obscures
the crossovers and thus increases the uncertainties in values of the inflection points.

For each level of gradient flow, the values of $ \mu_E $ and $ \mu_B $ differ just outside the
$1\sigma$ level for all the sweep numbers considered. This slight deviation may be a finite
temperature effect associated with the different spatial and temporal extents of the lattice and it
will be important to examine this further by considering lattices with a larger temporal extent.

In light of the chemical potentials available to us at this stage of our analysis, we are content
that the sigmoid function fits support a transition at a chemical potential near the anticipated
value of $am_\pi / 2 = 0.323(4)$ \cite{Cotter:2012mb}.

\begin{table}[tb]
\caption{Parameters $ \mu_E $ and $ \mu_B $ describe the point of inflection $ \mu_c $ in the
  squared chromoelectric and chromomagnetic field strengths as defined in Eq.~(\ref{eq: sig}).  }
\label{tab:sig}
\centering
 \begin{ruledtabular}
 \begin{tabular}{c c c c} 
 Gradient flow & Sweeps & $ a\mu_E $ & $ a\mu_B $ \\
 \hline
 \noalign{\smallskip}
 \multirow{3}{4em}{Moran} & 200 & 0.301 $\pm$ 0.002 & 0.295 $\pm$ 0.003 \\ 
 & 400 & 0.300 $\pm$ 0.002 & 0.290 $\pm$ 0.004 \\
 & 600 & 0.297 $\pm$ 0.003 & 0.286 $\pm$ 0.006 \\
 \hline
 \noalign{\smallskip}
 \multirow{4}{4em}{Iwasaki} & 150 & 0.302 $\pm$ 0.002 & 0.295 $\pm$ 0.003 \\ 
 & 300 & 0.302 $\pm$ 0.003 & 0.293 $\pm$ 0.004 \\
 & 450 & 0.302 $\pm$ 0.004 & 0.292 $\pm$ 0.005 \\
 & 600 & 0.302 $\pm$ 0.005 & 0.293 $\pm$ 0.006 \\
 \end{tabular}
 \end{ruledtabular}
\end{table}

In order to better constrain our estimate of the critical chemical potential of the phase
transition, we have generated several new ensembles in the region of the finite-volume crossover.
Each $16^3 \times 24$ ensemble is composed of approximately 200 gauge field configurations, with
$\beta=1.9$, $\kappa=0.1680$, and $a j = 0.04$. We introduce the new values $a \mu=
0.28,\ 0.29,\ 0.31,\ 0.32,\ 0.33,\ 0.34$ and $0.35$.  Including our earlier ensembles at
$a\mu=0.30$ and 0.40, this gives us nine points in the transition region.

\begin{figure}[tb]
\centering
\includegraphics[width=\columnwidth]{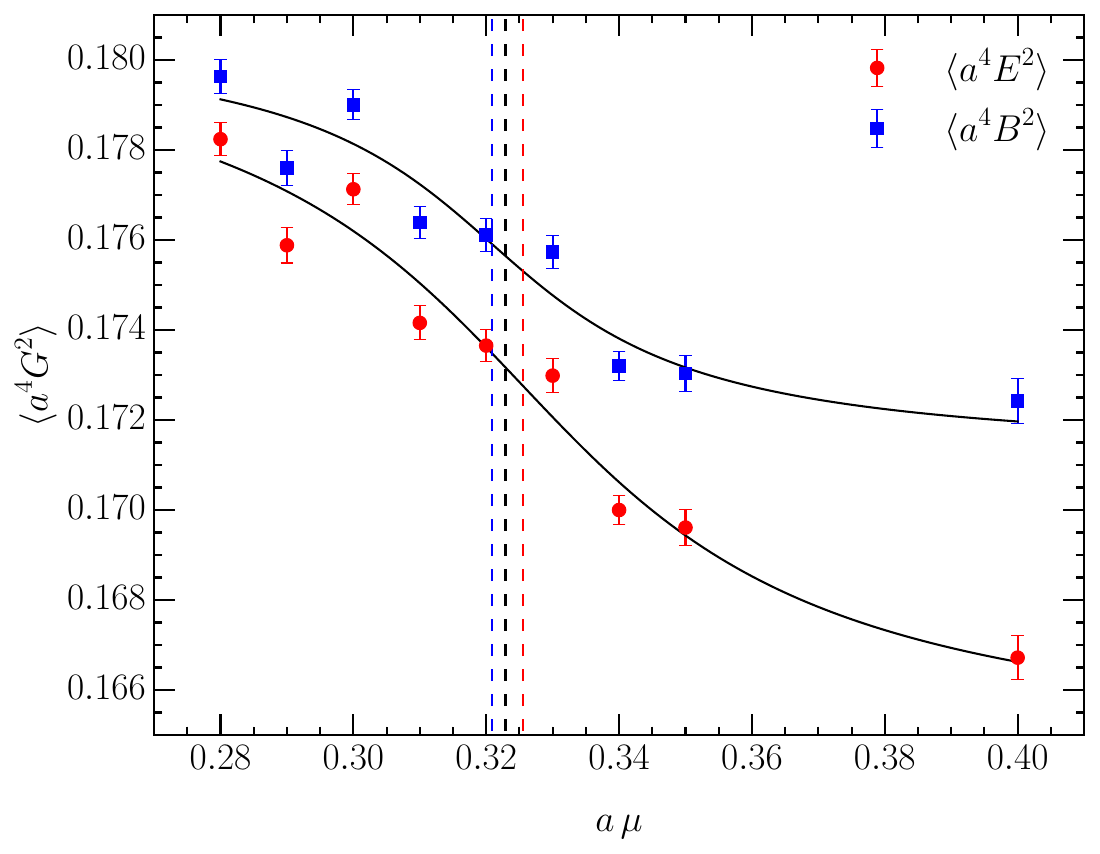}
\caption{Sigmoid fits to the squared chromoelectric, $\langle E^2 \rangle$, and chromomagnetic,
  $\langle B^2 \rangle$, field strengths obtained after Moran gradient flow at $t_g = 1.0$. This
  time all simulation results of the fit are in the crossover region.  The vertical dashed lines
  report the central values of the critical chemical potentials obtained from the vacuum field
  structure and the expected phase boundary at $m_\pi/2$ (black dashed line).}
\label{fig:CompSig}
\end{figure}

Figure \ref{fig:CompSig} presents our final determination of the critical chemical potential
obtained from changes in the ground-state field structure.  The critical values of the chemical
potentials obtained from fits to the $\langle E^2 \rangle$ and $\langle B^2 \rangle$ transitions
are
\begin{equation}
a \mu_E = 0.325(4) \mbox{\quad and\quad} a \mu_B = 0.321(4) \, ,
\end{equation}
respectively.  These values are both fully consistent with the anticipated phase boundary at $a
m_\pi /2 = 0.323(4)$.

\subsection{Difference in field strengths}
\label{diff}

\begin{figure}[tb]
\centering
\includegraphics[width=1.0\columnwidth]{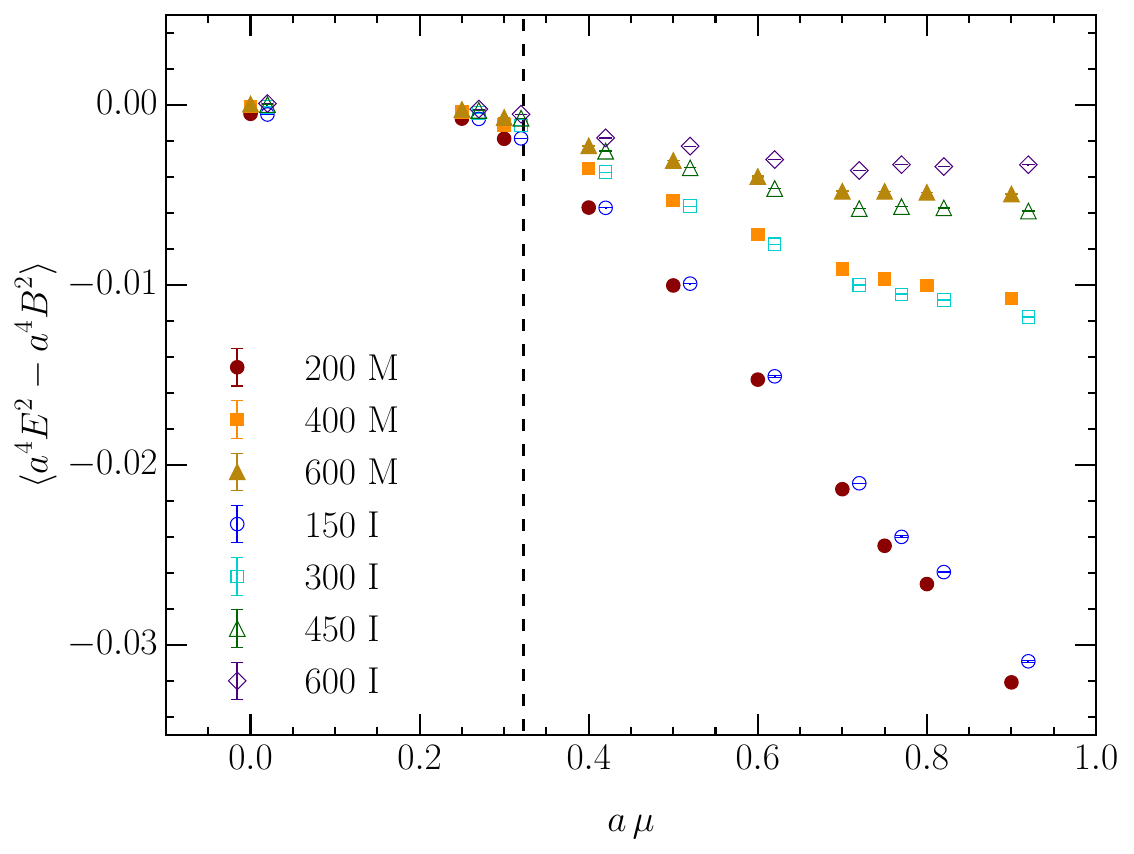}
\caption{Squared chromo-electromagnetic field strength difference, $\langle E^2 - B^2 \rangle$,
  is illustrated as a function of the chemical potential, $\mu$, for several different gradient
  flows.  Here, both Moran (M) and Iwasaki (I) gradient flows are reported with the number of
  gradient-flow sweeps at $\rho = 0.005$ indicated by the preceding integer in the legend.  }
\label{fig:EmB}
\end{figure}

\begin{figure}[tb]
\centering
\includegraphics[width=1.0\columnwidth]{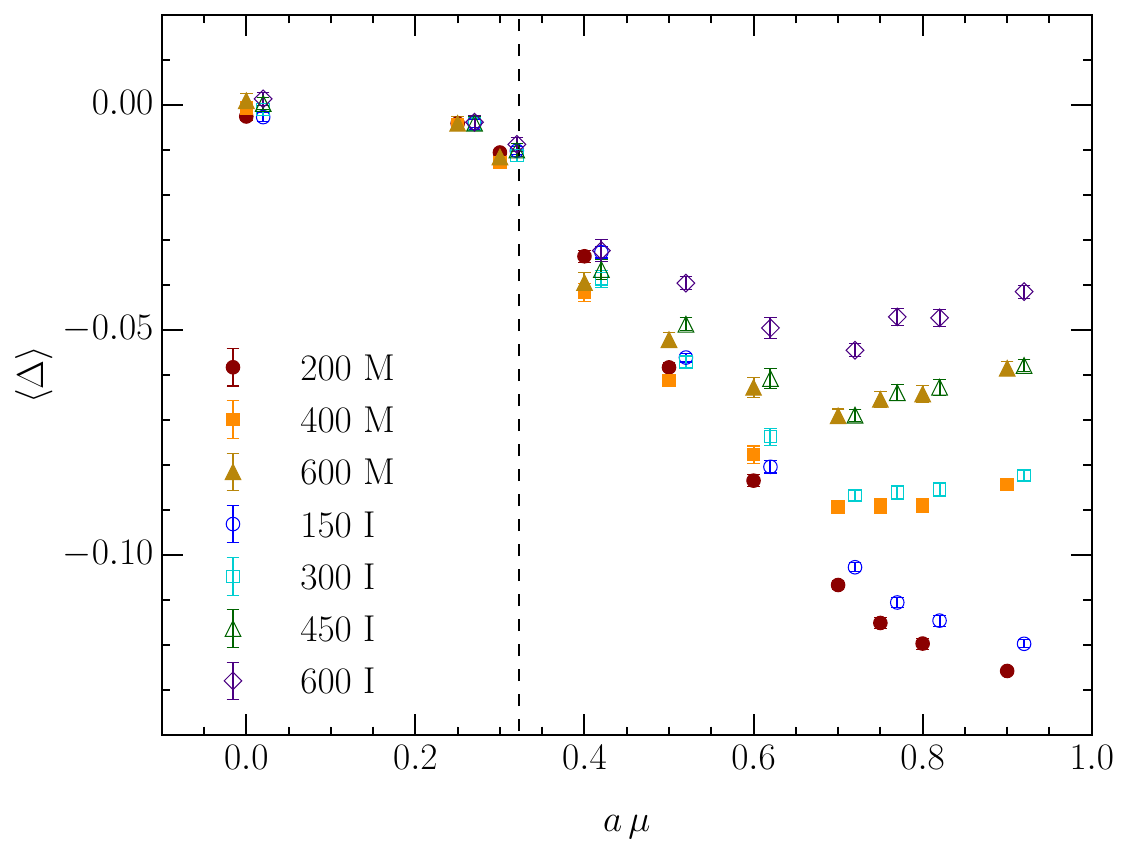}
\caption{Normalized chromo-electromagnetic field strength difference, $\langle \Delta
  \rangle$, of Eq.~(\ref{eq:delta}) is illustrated as a function of the chemical potential, $\mu$,
  for several different gradient flows.  As in Fig.~\ref{fig:EmB}, both Moran (M) and Iwasaki (I)
  gradient flows are reported with the number of gradient-flow sweeps at $\rho = 0.005$ indicated
  by the preceding integer in the legend.
}
\label{fig:Pct}
\end{figure}

The difference in the squared chromoelectric and the chromomagnetic field strengths further
illustrates the impact of the finite chemical potential. The dependence of this difference on the
gradient flow is illustrated in Fig.~\ref{fig:EmB}. Considering the earliest gradient flow times
first, the difference in squared chromoelectric and chromomagnetic field strengths for both Moran
and Iwasaki gradient flows increases monotonically as the chemical potential increases; the
difference increases at a lower rate for the moderate gradient flow, before becoming approximately
constant for higher numbers of sweeps. Again, lattice artifacts at the higher chemical potentials
may be affecting the results. Once again we see how large gradient flow times associated with
higher numbers of sweeps destroy the ground-state gauge field structures we aim to study.

To gain better intuition in quantifying the size of the effect, we introduce the normalized
difference
\begin{equation}
\label{eq:delta}
  \langle \Delta \rangle = \frac{\langle E^2 - B^2 \rangle}{\frac{1}{2} (\langle E^2 + B^2
    \rangle)}\, . 
\end{equation}

Figure~\ref{fig:Pct} shows the normalized difference plotted against the chemical potential. The
difference keeps increasing in magnitude as the chemical potential increases for the lowest number
of smoothing sweeps applied. However, the rate of magnitude increase in the difference drops at
higher chemical potentials, perhaps due to lattice artifacts influencing measurements at the
highest chemical potentials considered as we discussed previously.

Turning our attention to the results obtained following longer gradient flow trajectories, we see
how excess smoothing suppresses the difference even after it has been normalized. We conclude
that the process of smoothing is destroying the interesting signature of the effect of a finite
chemical potential on the ground-state field structure. It is also interesting to note the different
trend of the results at large chemical potentials under excess smearing.

In this regard, we recognize the importance of determining the minimum number of sweeps required to
reliably and quantitatively measure the chromo-electromagnetic field strengths. Unfortunately, the
quantities of interest evolve in a nontrivial manner with respect to flow time and a simple
extrapolation to zero flow time is insufficient.

In this light, we conclude that our best understanding of the modification of QC${}_2$D
ground-state fields in the presence of a finite chemical potential is obtained from the lowest
number of smearing sweeps consistent with stabilizing the topological charge. Moreover, we see that
the Moran 200 sweep results provide the maximum signature of ground-state field structure
modification. Given that further smoothing suppresses the signature, we regard the Moran 200 sweep
results as optimal for gaining insight into the modification of the ground-state field structure of
these ensembles.

\section{Conclusions}
\label{sec:conclusions}

We have presented the first examination of both qualitative and quantitative changes in the
ground-state field structure of QC${}_2$D at finite chemical potential.  Commencing with
visualizations of the subtle field structure changes, we were led to examine the differences in the
chromoelectric and chromomagnetic field strengths.

Drawing on the manner in which the chemical potential is introduced into the fermion action, it was
postulated that differences between the chromoelectric field strength and the chromomagnetic
field strength would manifest at large values of the chemical potential. Moreover, there was some
evidence that the strength of this difference would depend significantly on the methods used to
measure the field strengths on the lattice.

To address this problem, we performed a careful study of the amount of smearing required for the
lattice operators to provide an accurate indication of the field structure created through the
Monte Carlo procedure. Two ${\cal O}(a^4)$-improved measures of the topological charge density with
different higher-order corrections were introduced and their differences monitored as a function of
smearing. Four different gradient flow actions were considered to allow an estimation of the
systematic errors associated with the smoothing procedure. Dimensionless operators were created and
examined to understand the underlying features of these different smoothing actions and establish
smoothing thresholds for the earliest possible measurement of the chromoelectric and
chromomagnetic fields. Calibrations based on the action density confirmed the veracity of these
thresholds.

Examination of the chemical potential dependence of the squared field strengths confirms that minimal
smoothing provides a more robust signal. This emphasizes the importance of the quantitative
measures taken to determine the earliest possible gradient flow time at which one can reliably
calculate the ground-state field strengths.

The results show a significant gradient flow dependence. Qualitative differences in the chemical
potential dependence of the field strength measures were revealed. This prevents any attempts to
extrapolate back to zero gradient flow time.

\begin{figure}[tb]
\centering
\includegraphics[width=\columnwidth]{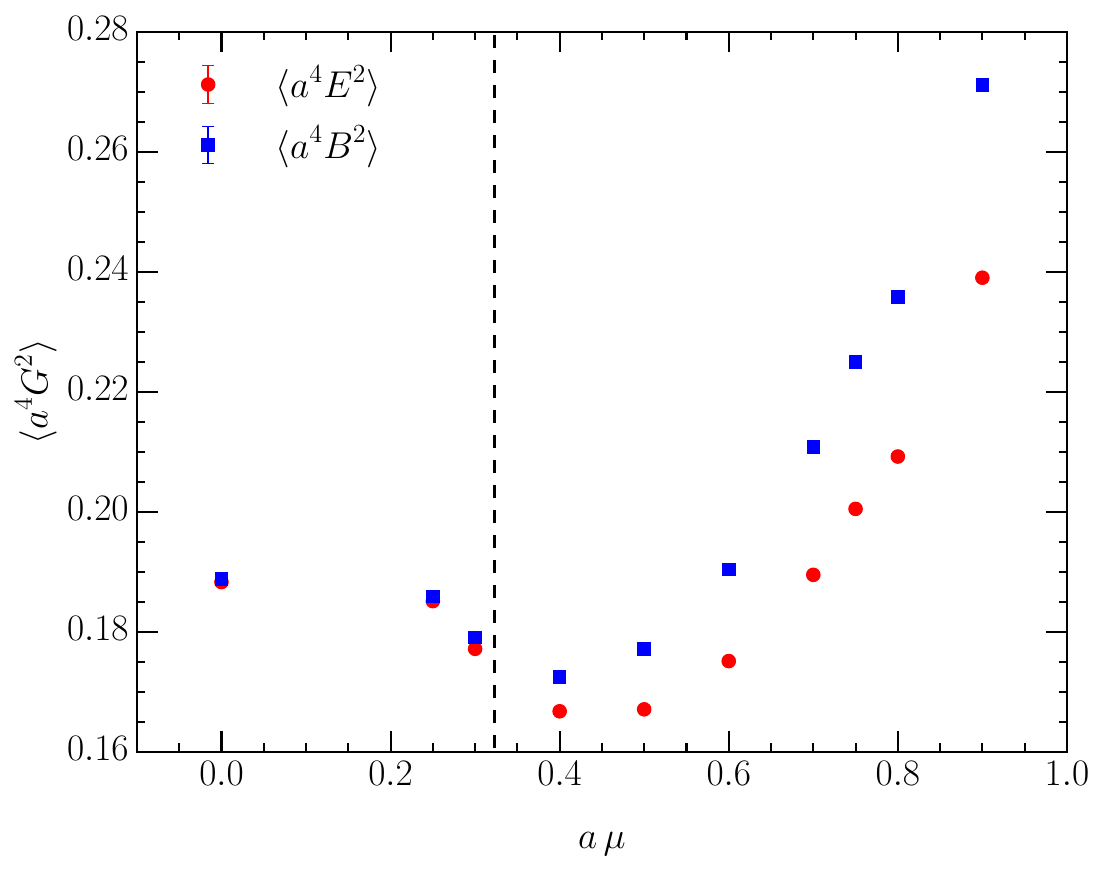}
\caption{Chemical potential dependence of the squared chromo-electromagnetic field strengths
  $\langle a^4E^2 \rangle$ and $\langle a^4B^2 \rangle$ at 200 sweeps of Moran gradient flow where
  the flow time $t_g = 1.0$.
}
\label{fig:Moran Gsq}
\end{figure}

\begin{figure}[tb]
\centering
\includegraphics[width=\columnwidth]{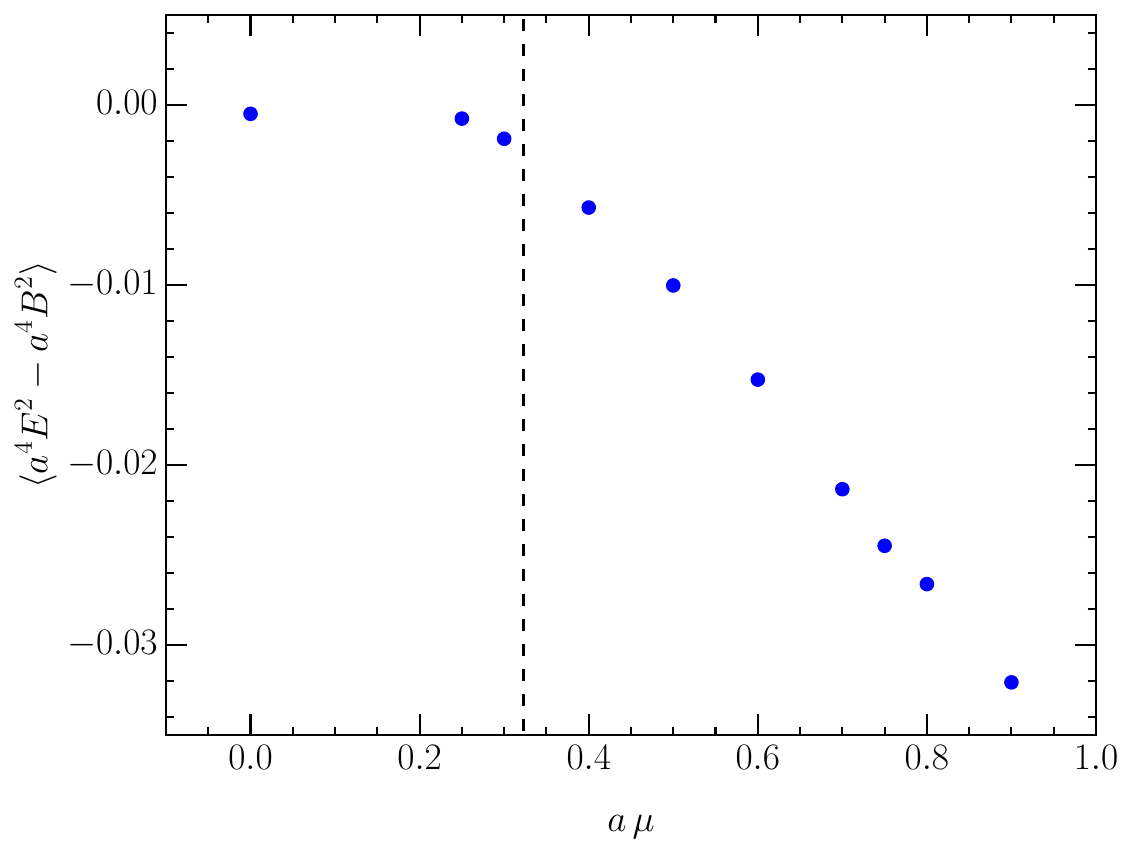}
\caption{Absolute difference of squared chromo-electromagnetic field strengths $\langle a^4E^2
  - a^4B^2 \rangle$ at 200 sweeps of Moran gradient flow where the flow time $t_g = 1.0$.  }
\label{fig:Moran EmB}
\end{figure}

\begin{figure}[tb]
\centering
\includegraphics[width=\columnwidth]{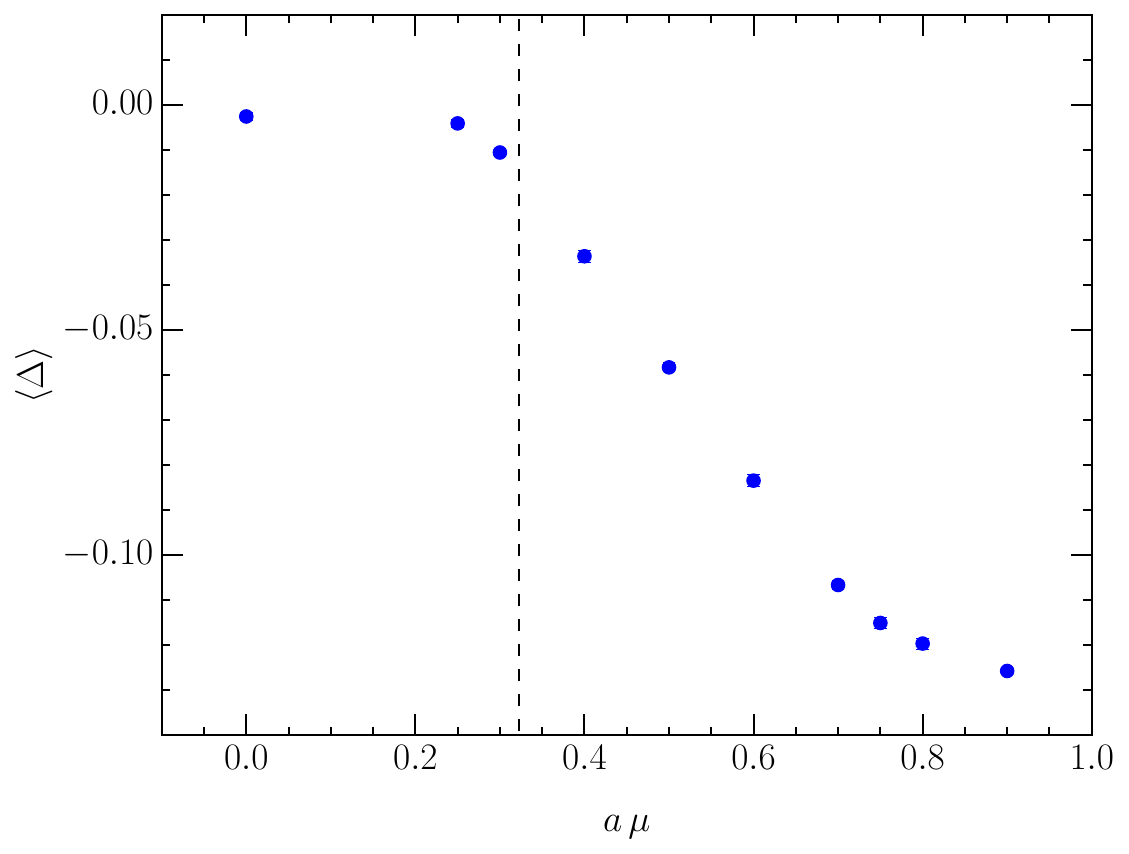}
\caption{Normalized difference of chromo-electromagnetic fields, $\langle \Delta \rangle$, of
  Eq.~(\ref{eq:delta}) at 200 sweeps of Moran gradient flow where the flow time $t_g = 1.0$.
}
\label{fig:Moran Pct}
\end{figure}

The Moran gradient flow action at 200 sweeps proved to provide the optimal signature of
ground-state field structure modification.  It is the earliest gradient flow time that ensures
controlled systematic errors in measures of the vacuum fields.  The overimproved Moran action was
designed to minimize structural changes during the gradient flow. It is therefore selected as the
best method for analyzing the density dependence of the ground-state SU(2) gauge field structures.
Hence, in summarizing our results, we present these results in Figs.~\ref{fig:Moran Gsq} through
\ref{fig:Moran Pct} as our best understanding of the modifications in ground-state field structures
due to a finite matter density.

The chemical potential dependence of the ground-state field strengths is
complicated. Figure~\ref{fig:Moran Gsq} shows a finite-volume crossover happening near $a\mu=0.3$
(the anticipated phase boundary), where both the chromoelectric and the chromomagnetic field
strengths are suppressed. Both the field strengths reach their respective minima at $a\mu=0.4$;
however, at higher chemical potentials, the ground-state field strengths restore and exceed their
zero chemical potential vacuum values.

The absolute squared field-strength difference, $\langle E^2 - B^2 \rangle$, is shown in
Fig.~\ref{fig:Moran EmB}. The point at $a\mu=0.0$ is $2.7 \, \sigma$ away from being zero; this is
likely a small finite temperature effect associated with the different finite temporal and spatial
lattice extents considered. Once again, the finite-volume crossover is observed near $a\mu=0.3$,
and the difference monotonically increases as the chemical potential increases, with the
chromomagnetic field strength being consistently larger than the chromoelectric field strength.

To gain intuition and quantify the magnitude of the effect, the normalized difference, $\langle
\Delta \rangle$, of Eq.~(\ref{eq:delta}) is constructed. This is illustrated in Fig.~\ref{fig:Moran
  Pct}. The difference at $a\mu=0.7$ is about $11\,\%$, which is a relatively small effect given
the significant differences in the physics encountered across the phase transition. The subtlety of
this effect is in accord with the early exploratory study of the distribution of topological charge
density at nonzero chemical potential \cite{Hands:2011hd} and the more recent studies of the
topological susceptibility \cite{Iida:2019rah,Iida:2024irv}.  The rate of increase in the
normalized difference drops after $a\mu=0.7$; however, this may be due to lattice artifacts.

We have also explored the position of the critical chemical potential of the phase
transition. Here, several new ensembles were generated in the region of the finite-volume
crossover.  Figure \ref{fig:CompSig} presents our final determinations. The critical values of
the chemical potentials obtained from fits to the $\langle E^2 \rangle$ and $\langle B^2 \rangle$
transitions are $a \mu_E = 0.325(4)$ and $a \mu_B = 0.321(4)$ respectively, and agree with the
anticipated phase boundary at $a m_\pi /2 = 0.323(4)$.

In summary, the chemical potential dependence of the QC${}_2$D ground-state field strengths is
subtle. On the other hand, the nature of the observables considered provide precise results on the
scale of these changes, such that clear conclusions may be drawn.

Our calculations have focused on observables that make the effect of the finite chemical potential
manifest.  Future quantitative calculations could explore quantities that have a more subtle
dependence on the chemical potential such as the distribution of topological charge density.

Avenues for further study include the development of even more sophisticated smoothing algorithms
that would allow one to study the field structure with even less smearing. The current calculations
are at the threshold of being able to reveal a source-term dependence in the measures
considered. Future work could provide high-statistics results that would enable one to address this
systematic error. Finally, it will be fascinating to examine the extent to which the phenomena
observed here extend to the three-color theory of QCD describing the strong force of nature.

\section*{Acknowledgments}

It is a pleasure to thank Ryan Bignell for his assistance in reading the modern file formats for
the finite-density gauge-field configurations investigated herein.
This work used the DiRAC Extreme Scaling service (Tursa) at the University of Edinburgh, managed by
the EPCC on behalf of the STFC DiRAC HPC Facility (www.dirac.ac.uk). The DiRAC service at Edinburgh
was funded by BEIS, UKRI and STFC capital funding and STFC operations grants. DiRAC is part of the
UKRI Digital Research Infrastructure.
D.L. acknowledges support from a John and Pat Hume doctoral scholarship.
This research was supported by the Australian Research Council through ARC Discovery Project Grant
DP210103706 (D.B.L.).

\section*{Data Availability}

The code used to generate the configurations studied herein is available via
Ref.~\cite{lawlor_2024_14201453}. 
The data presented and the associated plot scripts are publicly available in
Ref.~\cite{Zenodo.19324947}.

\bibliographystyle{utphys}
\bibliography{FiniteDensityVacuumRev}

\end{document}